\documentclass[12pt]{iopart}
\usepackage{iopams}  
\usepackage{graphicx}
\begin{document}

\title[Emergent Nesting of the Fermi Surface from Local-Moment Description]
{Emergent Nesting of the Fermi Surface from Local-Moment
Description of Iron-Pnictide High-$T_c$ Superconductors}


\author{J P Rodriguez$^1$, M A N Araujo$^{2,3}$, P D Sacramento$^3$}

\address{$^1$ Department of Physics and Astronomy, California State University at Los Angeles, Los Angeles, CA 90032}
\address {$^2$ Departamento de F\'{\i}sica, Universidade de \'Evora, P-7000-671, \'Evora,
Portugal}

\address {$^3$ CFIF, Instituto Superior T\'ecnico, Universidade de Lisboa,
 Av. Rovisco Pais, 1049-001 Lisboa, Portugal}
\ead{jrodrig@calstatela.edu}
\begin{abstract}
We uncover the low-energy spectrum of a $t$-$J$ model for electrons 
on a square lattice of spin-$1$ iron atoms
with $3d_{xz}$ and $3d_{yz}$ orbital character 
by applying
Schwinger-boson-slave-fermion mean-field theory and 
by exact diagonalization of one hole roaming over a $4\times 4\times 2$ lattice.
Hopping matrix elements are set to produce hole bands centered at zero two-dimensional (2D)
momentum in the free-electron limit.
Holes can propagate coherently in the $t$-$J$ model below a threshold Hund coupling
when long-range antiferromagnetic order across the
$d+ = 3d_{(x+iy)z}$ and $d- = 3d_{(x-iy)z}$
orbitals is established
by magnetic frustration that is off-diagonal in the orbital indices.
This leads to two hole-pocket Fermi surfaces centered at zero 2D momentum. 
Proximity to a commensurate spin-density wave (cSDW) 
that exists above the threshold Hund coupling
results in emergent Fermi surface pockets about cSDW momenta
at a quantum critical point (QCP).
This motivates the introduction of
a new Gutzwiller wavefunction for a cSDW metal state.
Study of the spin-fluctuation spectrum at cSDW momenta
indicates that the dispersion of the nested band of one-particle states 
that emerges is electron-type.
Increasing Hund coupling past the QCP can push the hole-pocket Fermi surfaces
centered at zero 2D momentum below the Fermi energy level,
in agreement with recent determinations of the
electronic structure of mono-layer iron-selenide superconductors.
\end{abstract}

\pacs{74.70.Xa, 75.10.Jm,75.30.Fv,75.30.Ds,71.10.Fd}
\maketitle

\section{Introduction}
The surprising discovery of high-temperature superconductivity in iron-pnictide compounds
is one of the
more recent unsolved puzzles in condensed matter physics\cite{new_sc}.
Determinations of the electronic structure in these superconductors
by angle-resolved photo-emission spectroscopy (ARPES) 
find hole bands that form  Fermi surface pockets at zero two-dimensional (2D) momentum
and electron bands that form Fermi surface pockets at 2D momenta
$(\pi / a) {\hat{\bi x}}$ and $(\pi / a) {\hat{\bi y}}$ \cite{zabolotnyy_09}\cite{fink09}\cite{brouet09}\cite{zhang11}.
Here $a$ denotes the lattice constant of the square lattice of iron atoms that
stacks up to form iron-pnictide materials.
Calculations of the electronic band structure that include all five iron $3d$ orbitals,
but that assume only weak inter-electron repulsion,
are consistent with the ARPES results\cite{singh_du_08}\cite{dong_08}\cite{graser_09}.
By contrast,
the simplest tight-binding model for the electronic structure of iron-pnictide materials
that include only the $3d_{xz}$ and $3d_{yz}$ orbitals 
can also produce such nested Fermi surface pockets\cite{raghu_08},
but it predicts a
Fermi surface pocket at 2D momentum $(\pi / a)({\hat{\bi x}} + {\hat{\bi y}})$
with a spectral weight that is much too strong\cite{graser_09}.
Iron-pnictide superconductors also have parent compounds that exhibit weak
commensurate spin-density-wave (cSDW) order at low temperature.
The ordered cSDW moment measured by elastic neutron diffraction
can reach values as low as $0.3$ Bohr magnetons ($\mu_B$) \cite{delacruz}.
By comparison, band structure calculations
predict an ordered magnetic moment of $2 \, \mu_B$ that is much larger.
Frustrated Heisenberg models that assume local magnetic moments at each iron atom
can successfully account for the
weak cSDW that exists in parent compounds\cite{Si&A}\cite{jpr_ehr_09}\cite{thalmeier_10}\cite{jpr10},
on the other hand.
They can also give a good account of the low-energy spin excitations near  cSDW momenta
that have been uncovered in iron-pnictide systems 
by inelastic neutron scattering\cite{zhao_09}\cite{diallo_09}\cite{hayden_10}\cite{Park_10}\cite{Li_10}\cite{liu_12}.
Such Heisenberg models have an insulating groundstate, however,
that is a result of strong inter-electron repulsion.
This fact conflicts with the metallic nature of iron-pnictide superconductors 
and their parent compounds.



We identify a way to resolve this dilemma from the limit of strong inter-electron repulsion
by injecting a low concentration of  mobile hole charges into a local-moment cSDW\cite{jpr_ehr_09}.
Spin-1/2 electrons are localized on
$d+ = 3d_{(x + iy)z}$  and $d- = 3d_{(x - iy)z}$ orbitals of each iron site,
and they can hop to unoccupied orbitals in neighboring iron atoms.
The $d{\pm}$ orbital basis maximizes the Hund's Rule coupling,
which therefore maximizes the tendency to form local magnetic moments per iron atom.
Hopping matrix elements are chosen
to produce 2D hole bands centered at zero momentum.
As Hund's Rule coupling weakens,
both Schwinger-boson-slave-fermion mean-field theory
and  exact computer calculations for one hole that roams over a $4\times 4$ lattice
find evidence for a quantum phase transition
into a hidden half-metal state
that shows long-range antiferromagnetic order across the $d+$ and $d-$ orbitals\cite{jpr11}.
Coherent intra-orbital  hole motion persists because of the hidden magnetic order.
This  yields two Fermi surface hole pockets centered at zero 2D momentum.
Nested Fermi surface pockets emerge at cSDW momenta as Hund coupling
increases past the quantum critical point.
(See Fig. \ref{conjecture_a}.)
In particular,
the exact calculations find
nested one-particle states at 
wavenumbers $(\pi / a){\hat{\bi x}}$ and $(\pi / a){\hat{\bi y}}$
that have  predominantly $3d_{yz}$ and  $3d_{xz}$ orbital character, respectively.
(See Fig. \ref{spctrm_cp_a}.)
This inspires us to introduce
a new Gutzwiller wavefunction\cite{Gutz} in section 5
that exhibits a low carrier density and cSDW nesting.
Further,
study of the spectrum for cSDW spin fluctuations in the present local-moment description
argues strongly for electron-type dispersion of the emergent Fermi surface pockets.

The local-moment description that we use to describe low-energy electronic physics in iron
superconductors makes two additional predictions.  
First, zero-energy hidden spin-wave excitations across the $d+$ and $d-$ orbitals
are present in the hidden half metal at the long wavelength limit.  
We show that these persist at low energy in the cSDW metal state
because of nesting between the two $d_{xz}$-$d_{yz}$ hole-pocket Fermi surfaces
that are centered at zero 2D momentum,
albeit with much reduced spectral weight.
The new Gutzwiller wavefunction introduced in section 5 is used to demonstrate this.
We point out that such low-energy hidden spin fluctuations are predicted by 
density functional theory (DFT) calculations
that exhibit  similar hole-pocket Fermi surfaces\cite{singh_du_08}\cite{dong_08}\cite{graser_09}.
Second, the fact that $d_{xz}$-$d_{yz}$ hole bands
lie below the Fermi surface in new interfacial
iron-selenide superconductors\cite{xue_12}\cite{zhou_12}\cite{zhou_13}\cite{peng_13}
is explained naturally
in our model
by moving off the quantum-critical point
that separates the hidden half metal from a cSDW metal.
Single-layer iron selenides have attained record critical temperatures 
within the class of iron superconductors.
DFT calculations, by comparison, typically yield that the $d_{xz}$-$d_{yz}$ hole bands
cross the Fermi surface in interfacial iron selenides\cite{B&C}.


\section{Two-Orbital t-J Model}
We shall now introduce a $t$-$J$ model that describes
the low-energy excitations of electrons
with $3d_{xz}$ and $3d_{yz}$ orbital character
in iron-pnictide high-temperature superconductors\cite{jpr11}.  
Spin-1/2 moments exist over the $3d_{(x+iy)z}$ and $3d_{(x-iy)z}$ orbitals at each iron atom.
These are the least localized orbitals within the 2D subspace spanned by the degenerate
$3d_{xz}$ and $3d_{yz}$ orbitals, and they therefore result in the largest 
possible Hund's Rule coupling constant, $-J_0$.
This fact is demonstrated explicitly in Appendix A
for the case of hydrogenic $3d$ orbitals with the bare Coulomb interaction.
In contrast, the commonly used $3d_{xz}$ and $3d_{yz}$ orbitals from Chemistry
are the most localized ones within the same 2D subspace\cite{f-b_60}\cite{e-r_63}\cite{Fulde}, 
and they result in the smallest possible Hund's Rule coupling.
The isotropic nature of the $3d_{(x \pm iy)z}$ orbitals also implies
isotropic Heisenberg exchange coupling constants $J_1$ and $J_2$
across nearest-neighbor and next-nearest-neighbor links of the
square lattice of iron atoms within each iron-pnictide layer.
The two-orbital $t$-$J$ model over a square-lattice of iron atoms in an isolated layer then reads
\begin{eqnarray}
H = &-\sum_{\langle i,j \rangle} \sum_{\alpha , \beta} \sum_{s} 
(t_1^{\alpha,\beta} c_{i, \alpha,s}^{\dagger} c_{j,\beta,s} + {\rm h.c.})
     -\sum_{\langle\langle i,j \rangle\rangle} \sum_{\alpha , \beta} \sum_{s} 
(t_2^{\alpha,\beta} c_{i, \alpha,s}^{\dagger} c_{j,\beta,s} + {\rm h.c.}) \nonumber \\
    &+{1\over 2} J_0 \sum_i \biggl[\sum_{\alpha} {\bi S}_{i, \alpha}\biggr]^2 +
    \sum_{\langle i,j \rangle} \sum_{\alpha , \beta} J_1^{\alpha,\beta}
                                      {\bi S}_{i, \alpha} \cdot {\bi S}_{j, \beta} +
    \sum_{\langle\langle i,j \rangle\rangle} \sum_{\alpha , \beta} J_2^{\alpha,\beta}
                                      {\bi S}_{i, \alpha} \cdot {\bi S}_{j, \beta} \nonumber \\
   &+{\rm lim}_{U_0\rightarrow\infty}\sum_i \sum_{\alpha} U_0 n_{i,\alpha,\uparrow} n_{i,\alpha,\downarrow}
   + 
\sum_i U_0^{\prime} n_{i,d+} n_{i,d-}. \nonumber \\
\label{tJ}
\end{eqnarray}
Above, ${\bi S}_{i,\alpha}$ is the spin operator that acts on the spin $s_0 = 1/2$ state
localized on orbital $\alpha$ at site $i$,
while $c_{i, \alpha,s}^{\dagger}$ is the corresponding electron creation operator.
The orbitals $\alpha$ are 
either $d+ = 3d_{(x+iy)z}$ or $d- = 3d_{(x-iy)z}$ ,
while the sites $i$ run over the square lattice of iron atoms.  
The constraint against double-occupancy at a site-orbital
is enforced by the divergent  Hubbard term, 
where $n_{i,\alpha,s} = c_{i, \alpha,s}^{\dagger} c_{i, \alpha,s}$ 
is the occupation operator. 
Also, the last term above gives the energy cost for a pair of holes at an iron site, $U_0^{\prime} > 0$,
where $n_{i,\alpha} = n_{i,\alpha,\uparrow} + n_{i,\alpha,\downarrow}$.
The isotropy and the degeneracy of the $d+$ and $d-$ orbital states
yields two independent and isotropic Heisenberg exchange coupling constants
for nearest-neighbor and for next-nearest-neighbor links,
$\langle i,j \rangle$ and $\langle\langle i,j \rangle\rangle$,
respectively, $n = 1$ and $2$:
\begin{equation}
J_{n}^{d+,d+} = J_{n}^{\parallel} = J_{n}^{d-,d-}
\quad {\rm and} \quad
J_{n}^{d+,d-} = J_{n}^{\perp} = J_{n}^{d-,d+}.
\label{jays}
\end{equation}
Correlated hopping of an electron in orbital $\beta$
to a neighboring unoccupied
orbital $\alpha$
is controlled by the
hopping matrix elements $t_{1}^{\alpha,\beta}$ and $t_{2}^{\alpha,\beta}$.
Let us take real hopping matrix elements in the $(3d_{xz} , 3d_{yz})$ orbital basis\cite{raghu_08}.
The hopping matrix elements in the present $d\pm$ basis are then given by
\numparts
\begin{eqnarray}
t_{n}^{d\pm,d\pm} = {1\over 2}[t_{n}^{x,x} + t_{n}^{y,y}]
\pm {i\over 2} [t_{n}^{x,y} - t_{n}^{y,x}], \\
t_{n}^{d\pm,d\mp} = {1\over 2}[t_{n}^{x,x} - t_{n}^{y,y}]
\mp {i\over 2} [t_{n}^{x,y} + t_{n}^{y,x}]. 
\end{eqnarray}
\endnumparts
The symmetry relation $t_{1(2)}^{x,y} = t_{1(2)}^{y,x}$
in the $(3d_{xz} , 3d_{yz})$ orbital basis
then yields diagonal hopping matrix elements that are real and isotropic
in the $(d+$, $d-)$ basis:
\numparts
\begin{eqnarray}
\qquad t_{1}^{d\pm,d\pm}(\hat{\bi x}) & =  t_{1}^{d\pm,d\pm}(\hat{\bi y}) & \quad{\rm and}\quad {\rm Im}\, t_{1}^{d\pm,d\pm} = 0, \\
t_{2}^{d\pm,d\pm}({\hat\bi x}+{\hat\bi y}) & =  t_{2}^{d\pm,d\pm}({\hat{\bi y}}-\hat{\bi x}) & \quad{\rm and}\quad {\rm Im}\, t_{2}^{d\pm,d\pm} = 0.
\end{eqnarray}
\endnumparts
The off-diagonal hopping matrix elements $t_{n}^{d\pm,d\mp}$ have $d$-wave symmetry, on the other hand.  
Also, the identities 
$t_1^{x,y} = 0 = t_1^{y,x}$ yields real off-diagonal hopping matrix elements 
across nearest neighbors, while
the symmetry relation $t_2^{x,x} = t_2^{y,y}$ yields
pure-imaginary off-diagonal hopping matrix elements across next-nearest neighbors:
\numparts
\begin{eqnarray}
\qquad t_1^{d\pm,d\mp} (\hat{\bi x}) & = - t_1^{d\pm,d\mp} (\hat{\bi y}) & \quad{\rm and}\quad {\rm Im}\, t_{1}^{d\pm,d\mp} = 0, \\
t_2^{d\pm,d\mp} (\hat{\bi x}+\hat{\bi y}) & = - t_2^{d\pm,d\mp} (\hat{\bi y}-\hat{\bi x}) &\quad {\rm and}\quad {\rm Re}\, t_{2}^{d\pm,d\mp} = 0.
\end{eqnarray}
\endnumparts
Henceforth, we shall change the notation for the hopping matrix elements to
$t_{1(2)}^{\parallel} = t_{1(2)}^{d\pm,d\pm}$, to
$t_{1}^{\perp} = t_{1}^{d\pm,d\mp}$, and to
$-t_{2}^{d-,d+} = t_{2}^{\perp} = t_{2}^{d+,d-}$.

Last, it is useful to point out that because $t_1^{\perp}$ is real,
a global swap of the orbitals, 
$d\pm\rightarrow d\mp$,
is an exact symmetry of the two-orbital $t$-$J$ model (\ref{tJ}) 
in the absence of next-nearest neighbor inter-orbital hopping: $t_2^{\perp} = 0$.
Let $P_{d\bar d}$ denote the global swap operation of the $d+$ and $d-$ orbitals. 
Eigenstates of the $t$-$J$ model Hamiltonian (\ref{tJ}) are then even ($+$) or  odd ($-$) under it,
with respective forms $|\Psi\rangle \pm P_{d\bar d}|\Psi\rangle$.
Further, in the limit $J_0\rightarrow -\infty$ where Hund's Rule is obeyed, 
a spin triplet state exists at iron atoms that do not have any holes.  
Such spin-$1$ local moments are clearly even under $P_{d\bar d}$.
This implies that a one-hole state that is even under $P_{d\bar d}$ is in a local $3d_{xz}$ orbital
when Hund's Rule is obeyed,
and that a one-hole state that is odd under $P_{d\bar d}$ is in a local $3d_{yz}$ orbital.
At the opposite extreme $J_0\rightarrow +\infty$ where Hund's Rule is maximally violated,
a spin singlet state exists at iron atoms without any holes.    
Such spin-$0$ iron sites are clearly odd under $P_{d\bar d}$.
In the case of an even number of iron sites,
this implies that a one-hole state that is even under $P_{d\bar d}$ instead
is in a local $3d_{yz}$ orbital,
and that a one-hole state that is odd under $P_{d\bar d}$ instead
is in a local $3d_{xz}$ orbital!
One-hole states that are even or odd under $P_{d\bar d}$ must then have a mixture of
$3d_{xz}$ and $3d_{yz}$ orbital character at finite Hund's Rule coupling by continuity.

And because $t_2^{\perp}$ is pure imaginary, the global orbital swap operation
is an exact symmetry in the absence of nearest-neighbor inter-orbital hopping, 
$t_1^{\perp} = 0$,
after making the global gauge transformation of the orbitals
$|d\pm\rangle \rightarrow i^{\pm 1/2} |d\pm\rangle$.
Eigenstates of the $t$-$J$ model Hamiltonian (\ref{tJ}) 
are then even ($+$) or  odd ($-$) under
the combined operation $P_{d\bar d}^{\prime}$,
where they have the form $|\Psi\rangle \pm P_{d\bar d}^{\prime}|\Psi\rangle$.
Notice that the global gauge transformation rotates the orbital axes by 45 degrees.
Repeating the previous arguments yields that
a one-hole state that is even under $P_{d\bar d}^{\prime}$ is in a local $3d_{x^{\prime}z}$ orbital
when Hund's Rule is obeyed,
and that a one-hole state that is odd under $P_{d\bar d}^{\prime}$ 
is in a local $3d_{y^{\prime}z}$ orbital in such case. 
Here $(x^{\prime} , y^{\prime})$ denote the planar orbital coordinates
along the next-nearest-neighbor links.

It is possible that iron-pnictide materials show low-energy electronic excitations with
orbital character other than $3d_{xz}$ and $3d_{yz}$.
DFT calculations predict that certain
Fermi surfaces have $3d_{xy}$ orbital character, for example\cite{graser_09}.
We work within the approximation that the Hamiltonian  that describes
the remaining set of orbitals commutes with the two-orbital $t$-$J$ model (\ref{tJ}).
In more physical terms, we therefore assume that all Fermi surfaces
with $3d_{xz}$ or $3d_{yz}$ character have no other orbital content.
A recent report of
electronic structure in
optimally doped Ba(Fe$_{1-x}$Co$_x$)$_2$As$_2$
from ARPES 
is consistent with this restriction \cite{zhang11}.


\section{Hidden Half Metal State}
In general, the Heisenberg exchange constants have a direct ferromagnetic contribution
from the exchange Coulomb integral 
and an indirect antiferromagnetic contribution
from the super-exchange mechanism through the pnictide atom\cite{Si&A}\cite{anderson_50}:
\begin{equation}
J_{1(2)}^{\alpha,\beta} =
J_{1(2)}^{\alpha,\beta} ({\rm drct}) + J_{1(2)}^{\alpha,\beta} ({\rm sprx}).
\label{drct-sprx}
\end{equation}
We shall assume that the super-exchange contribution 
is independent of the orbital indices:
$J_{1(2)}^{\alpha,\beta} ({\rm sprx}) = J_{1(2)}^{({\rm sprx})}$.
This is the case if the pnictide orbital in question is
the $p_x$, $p_y$, or $p_z$ one, for example.
Recent density-functional theory calculations for the electronic structure of iron-pnictide
materials find that direct exchange and super exchange can cancel out across nearest
neighbors\cite{ma_08}\cite{lu_11}.
More generally,
we shall assume that $J_{1(2)}^{d\pm, d\mp} ({\rm drct})$
is much  smaller in magnitude than  $J_{1(2)}^{d\pm, d\pm} ({\rm drct})$.
This assumption is borne out in the limit of large overlap between
neighboring iron $3d$ orbitals,
where $-J_{1(2)}^{d\pm, d\mp} ({\rm drct})$
is $10$ times smaller than $-J_{1(2)}^{d\pm, d\pm} ({\rm drct})$.
[See Appendix A, Eq. (\ref{ratio}).] 
We are then left with
a net antiferromagnetic Heisenberg exchange (\ref{drct-sprx}) across nearest neighbors
between different orbitals due to super-exchange, $J_1^{\perp} > 0$, 
and with a net nearest-neighbor
Heisenberg exchange (\ref{drct-sprx}) across the same orbital 
that is weaker, and possibly even ferromagnetic,
$J_1^{\parallel} < J_1^{\perp}$.
Last, DFT calculations find that
$J_{2}^{\alpha,\beta} ({\rm drct})$ is negligible\cite{ma_08}. 
This leaves antiferromagnetic Heisenberg exchange across next-nearest neighbors
due to super-exchange, which we have assumed to be invariant with respect
to a rotation of the orbital basis:
$J_2^{\parallel} = J_2^{\perp} > 0$.

\begin{figure}
\includegraphics{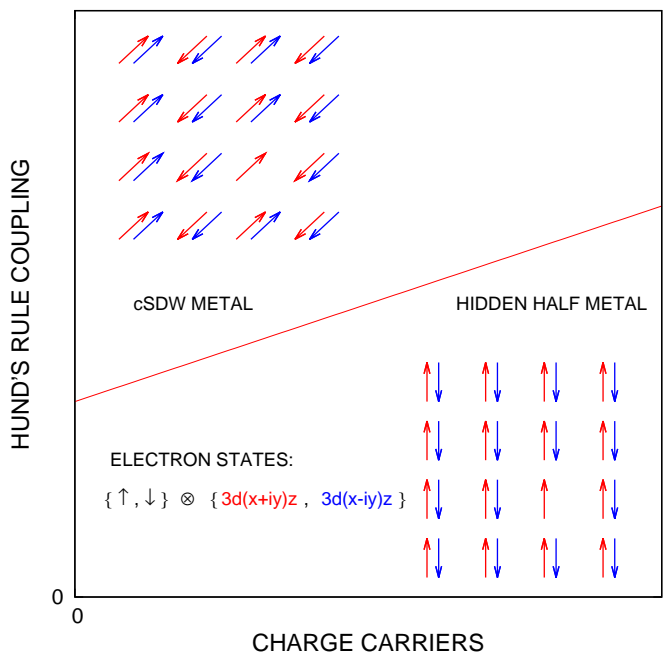}
\caption{Shown is the phase diagram obtained from the Schwinger-boson-slave-fermion mean-field
theory of the two-orbital $t$-$J$ model, Eq. (\ref{tJ}).
The red line marks a quantum-critical point that separates a charge-carrier poor cSDW metal from
a charge-carrier rich hidden half metal at fixed Hund's Rule coupling.}
\label{phase_diagram}
\end{figure}

In the absence of mobile holes,
the off-diagonal magnetic frustration described
by the above list of Heisenberg exchange constants 
results in a commensurate spin-density wave (cSDW) along
one of the principal axes of the square lattice of iron atoms
at strong Hund's Rule coupling, $-J_0$,
and for $J_2 > J_1 /2$.
Here $J_{1(2)} = (J_{1(2)}^{\parallel} + J_{1(2)}^{\perp})/2$.
An antiferromagnetic state with ferromagnetic order over
the $d+$ and the $d-$ orbitals of opposite sign
appears at Hund's Rule coupling below a critical value of
$-J_{0c} = 2 (J_1^{\perp} - J_1^{\parallel}) - 4 J_2^{\parallel}$ 
in the large-$s_0$ limit\cite{jpr10}.
We call this state a hidden ferromagnet
because it displays no net ordered magnetic moment.
The critical Hund coupling, $-J_{0c}$, marks a quantum-critical phase transition between the cSDW at
strong Hund's Rule coupling and the hidden ferromagnet at weak Hund's Rule coupling.
(Cf. Fig. \ref{phase_diagram}.)
Observe now that injecting mobile holes while
inter-orbital hopping is suppressed 
preserves the $\nwarrow_{d+}\searrow_{d-}$ spin order of the hidden ferromagnetic
state in the classical limit\cite{jpr10}. 
This is a semi-classical picture of a hidden half-metal state with coherent
propagation of holes within each $d\pm$ orbital
that follows the dispersion relation
\begin{equation}
\varepsilon_e ({\bi k}) = -2t_1^{\parallel}(\cos\,k_x a + \cos\,k_y a)
-2t_2^{\parallel}(\cos\,k_+ a + \cos\,k_- a) ,
\end{equation}
where $k_{\pm} = k_x \pm k_y$.
It implies two degenerate hole Fermi surface pockets centered at zero momentum
when $t_1^{\parallel} < 0$ and $t_1^{\parallel} + 2 t_2^{\parallel} < 0$,
in qualitative agreement with ARPES on iron-pnictide superconductors.
Below, we shall demonstrate that the hidden half-metal state is robust
with respect to the addition of inter-orbital hopping.
We will also find emergent cSDW nesting at the quantum-critical
point that separates the hidden half-metal state 
from the cSDW state. It results in copies of the hole Fermi-surface pockets
centered at cSDW
wavenumbers $(\pi / a){\hat{\bi x}}$ and $(\pi / a){\hat{\bi y}}$ \cite{jpr11}.
We shall argue in section 5 that 
the dispersion of the latter is electron type.

%

\subsection{Schwinger-Boson-Slave-Fermion Mean-field Theory}
Unlike
the doped N\'eel state in the one-orbital nearest-neighbor $t$-$J$ model
for copper-oxygen planes in high-$T_c$ superconductors\cite{S_S_88}\cite{Trugman_88},
hole propagation in the hidden half metal does not produce strings of overturned spins
that disrupt the antiferromagnetic $\nwarrow_{d+}\searrow_{d-}$ order
when inter-orbital hopping is absent.
This suggests that the mean-field approximation
for the  Schwinger-boson-slave-fermion formulation of the two-orbital $t$-$J$ model (\ref{tJ})
can be applied to this state at weak inter-orbital hopping.
Within that approximation,
we show below that
the hidden half-metal state survives the introduction of inter-orbital hopping
by dynamically suppressing it.

The constraint against double occupancy 
in the $t$-$J$ model (\ref{tJ}) can be
enforced by replacing the electron annihilation operator
with a correlated electron annihilation operator 
$\tilde c_{i,\alpha,s} = b_{i,\alpha,s} f_{i,\alpha}^{\dagger}$
along with a new constraint
per site $i$, per orbital $\alpha$,
\begin{equation}
2 s_0 = b_{i,\alpha,\uparrow}^{\dagger} b_{i,\alpha,\uparrow}
+ b_{i,\alpha,\downarrow}^{\dagger} b_{i,\alpha,\downarrow}
+ f_{i,\alpha}^{\dagger} f_{i,\alpha},
\label{constraint}
\end{equation}
where $s_0 = 1/2$ is the electron spin \cite{kane89}\cite{Auerbach_Larson_91}.
Here, $b_{i,\alpha,\uparrow}$ and $b_{i,\alpha,\downarrow}$ are the annihilation operators for
a pair of Schwinger bosons,
while $f_{i,\alpha}$ is the annihilation operator for a spinless slave fermion.
The spin operator is then expressed in the usual way:
${\bi S}_{i,\alpha} =
{1\over 2}\hbar \sum_{s,s{\prime}}
f_{i,\alpha} b_{i,\alpha,s}^{\dagger}
{\bsigma}_{s,s^{\prime}} b_{i,\alpha,s^\prime} f_{i,\alpha}^{\dagger}$.
Henceforth, we shall average the dynamics of the slave fermions over the bulk by writing
${\bi S}_{i,\alpha} \cong (1-x) {1\over 2}\hbar \sum_{s,s{\prime}}
b_{i,\alpha,s}^{\dagger} {\bsigma}_{s,s^{\prime}} b_{i,\alpha,s^\prime}$
instead.  Here, $x$ denotes the concentration of holes per orbital.
We shall also henceforth neglect hole-hole repulsion at iron sites, $U_0^{\prime}$,
which should be a valid approximation at low  $x$. 
Following Arovas and Auerbach\cite{Arovas_Auerbach_88}\cite{Auerbach_bk}, 
we introduce symmetric versus antisymmetric mean fields
with respect to spin flip for the ferromagnetic versus the  antiferromagnetic links
of the hidden ferromagnetic state (Fig. \ref{phase_diagram}):
%
\begin{eqnarray}
& \langle  b_{i,d\pm,\downarrow}^{\dagger} b_{j,d\pm,\downarrow}\rangle 
 = Q_n^{\parallel} = \langle b_{i,d\pm,\uparrow}^{\dagger} b_{j,d\pm,\uparrow}\rangle
\qquad (n = 1, 2),
\label{Q_para} \\
\pm & \langle b_{i,d\pm,\downarrow}  b_{j,d\mp,\uparrow}\rangle 
 = Q_n^{\perp} = \mp\langle b_{i,d\pm,\uparrow} b_{j,d\mp,\downarrow}\rangle
\qquad (n = 0, 1, 2),
\label{Q_perp}
\end{eqnarray}
where $j=i$ for on-site links $n=0$,
where $j=i+\hat{\bi x}(\hat{\bi y})$ for nearest-neighbor links $n=1$, and
where $j=i\pm\hat{\bi x} + \hat{\bi y}$ for next-nearest-neighbor links $n=2$.
Next, we introduce meanfields associated with hopping by slave fermions:
\begin{eqnarray}
{1\over 2} \langle f_{i,d+}^{\dagger} f_{j,d+}\rangle = & P_n^{\parallel}  = 
{1\over 2} \langle f_{i,d-}^{\dagger} f_{j,d-}\rangle , \\
\label{P_para_1}
{1\over 2} \langle f_{i,d+}^{\dagger} f_{j,d-}\rangle^* = & P_n^{\perp}  = 
{1\over 2} \langle f_{i,d-}^{\dagger} f_{j,d+}\rangle
\qquad (n = 1, 2).
\label{P_perp_1}
\end{eqnarray}
Last, we introduce meanfields associated with hopping of the Schwinger bosons across different orbitals:
\begin{equation}
\langle b_{i,d+,s}^{\dagger} b_{j,d-,s}\rangle^*
= R_n^{\perp} = \langle b_{i,d-,s}^{\dagger} b_{j,d+,s}\rangle
\qquad (n = 1, 2),
\label{R_perp}
\end{equation}
where $s = \uparrow , \downarrow$.
Notice that all of the mean fields
show invariance under orbital swap.
Now assume that all of the mean fields are also homogeneous, and assume that the mean fields 
$P_{1(2)}^{\parallel}$, $Q_{1(2)}^{\parallel}$ and  $Q_{1(2)}^{\perp}$ are isotropic.
The mean-field approximation for the $t$-$J$ model Hamiltonian (\ref{tJ}) then has the form
$H_{MF} = H_0[P,Q,R] + H_b^{(+)} + H_b^{(-)} + H_f$,
where 
\begin{eqnarray*}
H_0 = & 2 \sum_i J_0^{\prime} Q_0^{\perp} Q_0^{\perp *} + \\
& 2\sum_{\langle i,j\rangle} \sum_{\alpha}
[-(2 t_1^{\parallel} Q_1^{\parallel} P_1^{\parallel} + {\rm c.c.})
- J_1^{\prime\parallel} Q_1^{\parallel} Q_1^{\parallel *}] + \\
& 2\sum_{\langle i,j\rangle} \sum_{\alpha}
[-(2 t_1^{\perp} R_1^{\perp} P_1^{\perp} + {\rm c.c.}) 
+ J_1^{\prime\perp} Q_1^{\perp} Q_1^{\perp *}] + \\
& 2\sum_{\langle\langle i,j\rangle\rangle} \sum_{\alpha}
[-(2 t_2^{\parallel} Q_2^{\parallel} P_2^{\parallel} + {\rm c.c.}) 
 - J_2^{\prime\parallel} Q_2^{\parallel} Q_2^{\parallel *}] + \\
& 2\sum_{\langle\langle i,j\rangle\rangle} \sum_{\alpha}
[-(2 t_2^{\perp} R_2^{\perp} P_2^{\perp} + {\rm c.c.}) 
+ J_2^{\prime\perp} Q_2^{\perp} Q_2^{\perp *}]
\end{eqnarray*}
consolidates the bilinear terms among the mean fields, where
\begin{equation*}
H_b^{(+)} = {1\over 2}\sum_{\bi k}
\left[ {\begin{array}{c}
b_{d+,\uparrow} \\ b_{d-,\uparrow} \\ {\bar b}_{d+,\downarrow}^{\dagger} \\ {\bar b}_{d-,\downarrow}^{\dagger}
\end{array} } \right]^{\dagger}
\left[ {\begin{array}{cccc}
\Omega_{\parallel}      & \Omega_{\perp}^{\prime *} & 0                         & +\Omega_{\perp}         \\
\Omega_{\perp}^{\prime} & \Omega_{\parallel}        & -\Omega_{\perp}           & 0                       \\
0                       & -\Omega_{\perp}           & \Omega_{\parallel}        & \Omega_{\perp}^{\prime} \\
+\Omega_{\perp}         & 0                         & \Omega_{\perp}^{\prime *} &\Omega_{\parallel}
\end{array} } \right]
\left[ {\begin{array}{c}
b_{d+,\uparrow} \\ b_{d-,\uparrow} \\ {\bar b}_{d+,\downarrow}^{\dagger} \\ {\bar b}_{d-,\downarrow}^{\dagger}
\end{array} } \right]
\end{equation*}
and
\begin{equation*}
H_b^{(-)} = {1\over 2}\sum_{\bi k}
\left[ {\begin{array}{c}
b_{d+,\downarrow} \\ b_{d-,\downarrow} \\ {\bar b}_{d+,\uparrow}^{\dagger} \\ {\bar b}_{d-,\uparrow}^{\dagger}
\end{array} } \right]^{\dagger}
\left[ {\begin{array}{cccc}
\Omega_{\parallel}      & \Omega_{\perp}^{\prime *} & 0                         & -\Omega_{\perp}         \\
\Omega_{\perp}^{\prime} & \Omega_{\parallel}        & +\Omega_{\perp}           & 0                       \\
0                       & +\Omega_{\perp}           & \Omega_{\parallel}        & \Omega_{\perp}^{\prime} \\
-\Omega_{\perp}         & 0                         & \Omega_{\perp}^{\prime *} &\Omega_{\parallel}
\end{array} } \right]
\left[ {\begin{array}{c}
b_{d+,\downarrow} \\ b_{d-,\downarrow} \\ {\bar b}_{d+,\uparrow}^{\dagger} \\ {\bar b}_{d-,\uparrow}^{\dagger}
\end{array} } \right]
\end{equation*}
are the pair of Hamiltonians for free Schwinger bosons, with matrix elements
\begin{eqnarray}
\Omega_{\parallel}({\bi k}) =
&\lambda 
+4\sum_{n=1,2}(J_n^{\prime\parallel} Q_n^{\parallel} + 2 t_n^{\parallel} P_n^{\parallel})\gamma_n({\bi k}),\\
\Omega_{\perp}({\bi k}) =
& \sum_{n=0,1,2} z_n J_n^{\prime\perp} Q_n^{\perp}\gamma_n({\bi k}),\\
\Omega_{\perp}^{\prime}({\bi k}) =
&4\sum_{n=x,y} t_1^{\perp}(\hat{\bi n}) P_1^{\perp}(\hat{\bi n}) \cos (k_n a) \nonumber \\
&+4\sum_{\pm} t_2^{\perp}(\hat{\bi y}\pm\hat{\bi x}) P_2^{\perp *}(\hat{\bi y}\pm\hat{\bi x}) \cos (k_{\pm} a),
\end{eqnarray}
and where
\begin{equation*}
H_f = \sum_{\bi k}
\left[ {\begin{array}{c}
f_{d+} \\ f_{d-}
\end{array} } \right]^{\dagger} 
\left[ {\begin{array}{cc}
\varepsilon_{\parallel}    & \varepsilon_{\perp}   \\
\varepsilon_{\perp}^*          & \varepsilon_{\parallel}
\end{array} } \right]
\left[ {\begin{array}{c}
f_{d+} \\ f_{d-}
\end{array} } \right]
\end{equation*}
%
is the Hamiltonian for free slave fermions,
with matrix elements
\begin{eqnarray}\varepsilon_{\parallel}({\bi k}) =
& 8\sum_{n=1,2} t_n^{\parallel} Q_n^{\parallel} \gamma_n({\bi k}),
\label{epsilon_para}\\
\varepsilon_{\perp}({\bi k}) =
& 4\sum_{n=x,y} t_1^{\perp}(\hat{\bi n}) R_1^{\perp}(\hat{\bi n}) \cos (k_n a) \nonumber \\
& +4\sum_{\pm} t_2^{\perp}(\hat{\bi y}\pm\hat{\bi x}) R_2^{\perp}(\hat{\bi y}\pm\hat{\bi x}) \cos (k_{\pm} a).
\label{epsilon_perp}
\end{eqnarray}
Above, the
destruction operators for Schwinger bosons of momentum $\pm\bi k$ are defined by
$b_{\alpha,s} ({\bi k}) = {N_{\rm Fe}}^{-1/2}\sum_i e^{-i{\bi k}\cdot{\bi r}_i} b_{i,\alpha,s}$ and
${\bar b}_{\alpha,s} ({\bi k}) = b_{\alpha,s} (-{\bi k})$,
while the destruction operator for a slave fermion of momentum $\bi k$ is
$f_{\alpha} ({\bi k}) = {N_{\rm Fe}}^{-1/2}\sum_i e^{-i{\bi k}\cdot{\bi r}_i} f_{i,\alpha}$.
Also above,  
$z_0 = 1$ and $z_{1(2)} = 4$ give the coordination number,
and we define
$\gamma_0({\bi k}) = 1$,
$\gamma_{1}({\bi k}) = {1\over 2} ({\rm cos}\, k_{x} a\,+\,{\rm cos}\, k_{y} a)$,
and $\gamma_2({\bi k}) = {1\over 2} ({\rm cos}\, k_+ a\,+\,{\rm cos}\, k_- a)$,
where $k_{\pm} = k_x \pm k_y$.
Last, $\lambda$ is the boson chemical potential
that enforces the constraint against double occupancy (\ref{constraint}) on {\it average}
over the bulk of the system,
and  the effect of mobile holes on the Heisenberg
spin-exchange is accounted for by
the effective exchange coupling  constants\cite{Auerbach_Larson_91}
$J^{\prime} = (1-x)^2 J$.

The spin-excitation spectrum of the hidden half-metal state
is obtained from the sum of the Schwinger-boson Hamiltonians,
$H_b^{(+)}+H_b^{(-)}$,
in two steps.
We first make a two-orbital {\it Bogoliubov} transformation of the boson field:
\numparts
\begin{eqnarray}
b_{\alpha,s} = &(\cosh \, \theta) \beta_{\alpha,s}
-({\rm sgn}\,\alpha)({\rm sgn}\, s) (\sinh \, \theta) \bar\beta_{\bar\alpha,\bar s}^{\dagger},
\label{Bgv_trns_a}\\
\bar b_{\bar\alpha,\bar s}^{\dagger} = &(\cosh \, \theta) \bar \beta_{\bar\alpha,\bar s}^{\dagger}
-({\rm sgn}\,\alpha)({\rm sgn}\, s)(\sinh \, \theta) \beta_{\alpha,s},
\label{Bgv_trns_b}
\end{eqnarray}
\endnumparts
%
with
$\cosh \, 2\theta = \Omega_{\parallel} / \omega_b^{(0)}$ and
$\sinh \, 2\theta =  \Omega_{\perp} / \omega_b^{(0)}$, where
$\omega_b^{(0)} = (\Omega_{\parallel}^2 - \Omega_{\perp}^2)^{1/2}$.
Here, we use the notation ${\bar d\pm} = d\mp$ in the orbital index
and $\bar s = -s$ in the spin index.
The sum
$H_b^{(+)} + H_b^{(-)}$ then transforms to
\begin{eqnarray*}
H_b = 
&{1\over 2}\sum_{\bi k} \sum_{s=\uparrow,\downarrow}
\left[ {\begin{array}{c}
\beta_{d+,s} \\ \beta_{d-,s} 
\end{array} } \right]^{\dagger}
\left[ {\begin{array}{cc}
\omega_b^{(0)}           & \Omega_{\perp}^{\prime *} \\
\Omega_{\perp}^{\prime}  & \omega_b^{(0)}            \\
\end{array} } \right]
\left[ {\begin{array}{c}
\beta_{d+,s} \\ \beta_{d-,s} 
\end{array} } \right] \nonumber \\
&+{1\over 2}\sum_{\bi k} \sum_{s=\uparrow,\downarrow}
\left[ {\begin{array}{c}
\bar\beta_{d+,s} \\ \bar\beta_{d-,s} 
\end{array} } \right]^{\rm T}
\left[ {\begin{array}{cc}
\omega_b^{(0)}           & \Omega_{\perp}^{\prime} \\
\Omega_{\perp}^{\prime *}  & \omega_b^{(0)}            \\
\end{array} } \right]
\left[ {\begin{array}{c}
\bar\beta_{d+,s}^{\dagger} \\ \bar\beta_{d-,s}^{\dagger}
\end{array} } \right].
\end{eqnarray*}
Let
$e^{2 i\delta_b({\bi k})}$ denote the phase factor of the 
inter-orbital matrix element $\Omega_{\perp}^{\prime}({\bi k})$,
and let
$k_0 = 0,\pi$ denote bonding $(+)$ and anti-bonding $(-)$ superpositions among the $d\pm$ orbitals
after making the gauge transformation $e^{\pm i\delta_b({\bi k})}$.
The similarity transform
\begin{equation}
\beta_{d\pm,s} ({\bi k}) = e^{\mp i\delta_b({\bi k})}     
[2^{-1/2}\beta_{s}(0,{\bi k})
\mp 2^{-1/2}\beta_{s}(\pi,{\bi k})]
\label{sim_trans_b}
\end{equation}
then reduces the Schwinger boson Hamiltonian to
\begin{equation}
H_b = {1\over 2}\sum_{k_0=0,\pi}\sum_{\bi k} \sum_{s=\uparrow,\downarrow}
\omega_b(k) [\beta_s^{\dagger} (k) \beta_s (k)
+ \beta_s (-k) \beta_s^{\dagger} (-k)],
\end{equation}
with the spectrum
\begin{equation}
\omega_b (k_0,{\bi k}) = [\Omega_{\parallel}^2({\bi k})-\Omega_{\perp}^2({\bi k})]^{1/2}
 + e^{i k_0} |\Omega_{\perp}^{\prime}({\bi k})|.
\label{w_b}
\end{equation}
The charge-excitation spectrum due to the slave fermions,
on the other hand, is obtained directly by the
similarity transform
\begin{equation}
f_{d\pm} ({\bi k}) = e^{\pm i\delta_f({\bi k})}
[2^{-1/2} f (0,{\bi k})
\mp 2^{-1/2} f (\pi,{\bi k})],
\label{rotate_f}
\end{equation}
where $e^{2 i \delta_f({\bi k})}$ denotes the phase factor of 
the inter-orbital matrix element $\varepsilon_{\perp}({\bi k})$.
It yields the diagonal form
\begin{equation}
H_f = \sum_{k_0=0,\pi}\sum_{\bi k} \varepsilon_f (k) f^{\dagger}(k) f(k)
\end{equation}
for the slave-fermion Hamiltonian, with spectrum
\begin{equation}
\varepsilon_f (k_0,{\bi k}) = \varepsilon_{\parallel}({\bi k}) + e^{i k_0} |\varepsilon_{\perp}({\bi k})|.
\label{e_f}
\end{equation}
The low-energy spin and charge excitations that result from $\omega_b$ and $\varepsilon_f$ 
will be discussed in detail below.

We must first obtain the mean fields $P$, $Q$ and $R$, however.
Taking the quantum-thermal average of the constraint against double occupancy (\ref{constraint}),
and averaging it over the bulk, we get the principal mean-field equation
\begin{equation}
s_0 + {1\over 2}(1-x) = {\cal N}^{-1} \sum_k (\cosh\, 2\theta) 
\Bigl({1\over 2} + n_B[\omega_b(k)]\Bigr),
\label{s_0_mf_eq}
\end{equation}
where ${\cal N} = 2 N_{\rm Fe}$, and where $n_B$ denotes the Bose-Einstein distribution:
$n_B(\omega) = [{\rm exp}(\omega/k_B T) - 1]^{-1}$.
Next, averaging the definitions of the mean fields (\ref{Q_para})-(\ref{R_perp}) over the bulk
yields self-consistent equations
\numparts
\begin{eqnarray}
Q_n^{\parallel} = {\cal N}^{-1} \sum_k \gamma_n({\bi k}) [\cosh\, 2\theta({\bi k})]
\Bigl({1\over 2} + n_B[\omega_b(k)]\Bigr) \qquad (n=1,2) 
\label{Q_para_mf_eq} \\
Q_n^{\perp} = {\cal N}^{-1} \sum_k \gamma_n({\bi k}) [\sinh\, 2\theta({\bi k})]
\Bigl({1\over 2} + n_B[\omega_b(k)]\Bigr) \qquad (n=0,1,2)
\label{Q_perp_mf_eq}
\end{eqnarray}
\endnumparts
and
\begin{eqnarray}
R_1^{\perp} (\hat{\bi n}) = {\cal N}^{-1} \sum_k (\cos\, k_n a)
e^{ik_0} e^{2i\delta_b({\bi k})} n_B[\omega_b(k)] \qquad (n = x,y) \nonumber \\
R_2^{\perp} (\hat{\bi y}\pm\hat{\bi x}) = {\cal N}^{-1} \sum_k (\cos\, k_{\pm} a)
e^{ik_0} e^{2i\delta_b({\bi k})} n_B[\omega_b(k)]
\label{R_perp_mf_eq}
\end{eqnarray}
associated with the Schwinger bosons,
and it yields self-consistent equations
\begin{equation}
P_n^{\parallel} = {1\over 2}{\cal N}^{-1} \sum_k \gamma_n({\bi k})
n_F[\varepsilon_f(k)] \qquad (n=1,2)
\label{P_para_mf_eq}
\end{equation}
and
\begin{eqnarray}
P_1^{\perp} (\hat{\bi n}) = {1\over 2} {\cal N}^{-1} \sum_k
(\cos\, k_n a) e^{ik_0} e^{2i\delta_f({\bi k})} n_F[\varepsilon_f(k)] \qquad (n = x,y) \nonumber \\
P_2^{\perp} (\hat{\bi y}\pm\hat{\bi x}) = {1\over 2} {\cal N}^{-1} \sum_k
(\cos\, k_{\pm} a) e^{ik_0} e^{2i\delta_f({\bi k})} n_F[\varepsilon_f(k)]
\label{P_perp_mf_eq_1}
\end{eqnarray}
associated with the slave fermions. 
Above, $n_F$ denotes the Fermi-Dirac distribution:
$n_F(\varepsilon) = ({\rm exp}[(\varepsilon-\mu)/k_B T] + 1)^{-1}$,
with a chemical potential $\mu$.
Let us now assume that the mean fields $R_1^{\perp}$ and $R_2^{\perp}$ are both isotropic, 
which will be shown to be self consistent.
This yields the form
\begin{eqnarray}
\varepsilon_{\perp} ({\bi k}) =
& 4 t_1^{\perp} ({\hat{\bi x}}) R_1^{\perp} [\cos(k_x a) - \cos(k_y a)] \nonumber \\
& + 4 t_2^{\perp}({\hat{\bi x}}+{\hat{\bi y}}) R_2^{\perp} [\cos(k_+ a) - \cos(k_- a)],
\label{epsilon_perp_bis}
\end{eqnarray}
for the inter-orbital matrix element experienced by slave fermions,
where $t_1^{\perp}$ is real and  where $t_2^{\perp}$ is pure imaginary.
It alternates in sign when ${\bi k}$ is rotated by $\pi/2$.
This $d$-wave symmetry implies that both
$P_1^{\perp}(\hat{\bi x}) + P_1^{\perp}(\hat{\bi y}) = 
{\cal N}^{-1} \sum_k e^{ik_0} e^{2i\delta_f({\bi k})} \gamma_1({\bi k}) n_F[\varepsilon_f(k)]$
and
$P_2^{\perp}(\hat{\bi x}+\hat{\bi y}) + P_2^{\perp}(\hat{\bi y}-\hat{\bi x}) = 
{\cal N}^{-1} \sum_k e^{ik_0} e^{2i\delta_f({\bi k})} \gamma_2({\bi k}) n_F[\varepsilon_f(k)]$
vanish.
Their mean-field equations therefore reduce to
\begin{eqnarray}
P_1^{\perp} (\hat{\bi x}) = {1\over 4} {\cal N}^{-1} \sum_k
e^{ik_0} e^{2i\delta_f({\bi k})} [(\cos\, k_x a)-(\cos\, k_y a)] n_F[\varepsilon_f(k)], \nonumber \\
P_2^{\perp} (\hat{\bi x} + \hat{\bi y}) = {1\over 4} {\cal N}^{-1} \sum_k
e^{ik_0} e^{2i\delta_f({\bi k})} [(\cos\, k_{+} a)-(\cos\, k_{-} a)] n_F[\varepsilon_f(k)],
\label{P_perp_mf_eq_2}
\end{eqnarray}
with opposite signs for the orthogonal links.  
The latter $d$-wave symmetry then yields the form
\begin{equation}
\Omega_{\perp}^{\prime}({\bi k}) =
8 t_1^{\perp}({\hat{\bi x}}) P_1^{\perp *}(\hat{\bi x}) \gamma_1({\bi k})
+ 8 t_2^{\perp}({\hat{\bi x}}+{\hat{\bi y}}) P_2^{\perp *}(\hat{\bi x}+\hat{\bi y}) \gamma_2({\bi k})
\label{Omega_perp_prime_1}
\end{equation}
for the inter-orbital matrix element experienced by Schwinger bosons.
It has $s$-wave symmetry, and it is thus consistent with the previous assumptions.

\begin{figure}
\includegraphics[scale=1.2]{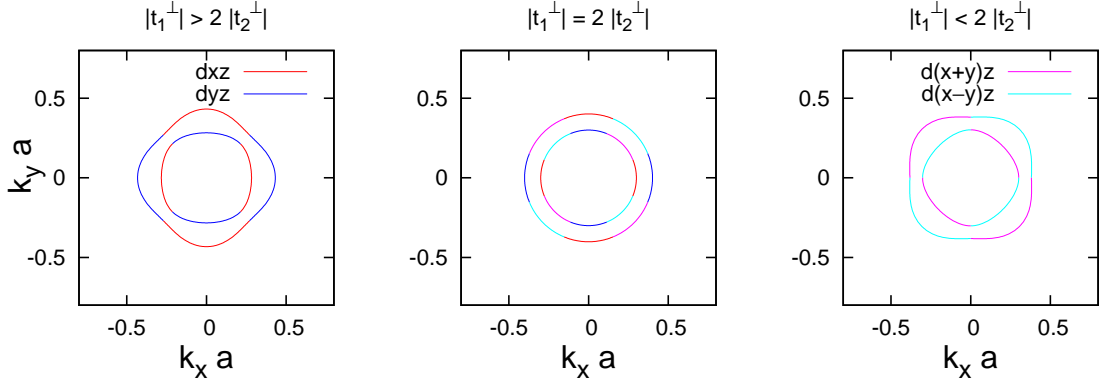}
\caption{Shown are the hole Fermi surfaces centered at zero 2D momentum:
$t_{\parallel} < 0$.  
The concentration of holes is universally set to $x = 0.01$, 
while the hopping matrix elements from left to right are
(a) $t_{\perp} = 0.40\, t_{\parallel}$ and $t_{\perp}^{\prime} = 0.23\, t_{\parallel}$ ,
(b) $t_{\perp} = 0.28\, t_{\parallel}$ and $t_{\perp}^{\prime} = 0.28\, t_{\parallel}$ , and
(c) $t_{\perp} = 0.23\, t_{\parallel}$ and $t_{\perp}^{\prime} = 0.40\, t_{\parallel}$ .
Above, orbital labels are only approximate near the points on the Fermi surfaces 
at which they change.}
\label{FS}
\end{figure}

To proceed further, we shall assume that both inter-orbital mean fields
for Schwinger bosons
$R_1^{\perp}$ and $R_2^{\perp}$ are real and positive.
This will also be confirmed self-consistently.
The slave-fermion matrix elements are then approximated by
\numparts
\begin{eqnarray}
\varepsilon_{\parallel}({\bi k}) = 
&\varepsilon_{\parallel}(0) - t_{\parallel} |{\bi k}|^2 a^2
\label{e_f_para} \\
\varepsilon_{\perp} ({\bi k}) =
&-[t_{\perp} \cos (2\phi) +  i t_{\perp}^{\prime} \sin (2\phi)] |{\bi k}|^2 a^2
\label{e_f_perp}
\end{eqnarray}
\endnumparts
for ${\bi k}$ near zero, where
$t_{\parallel} = 2 t_1^{\parallel} Q_1^{\parallel} + 4 t_2^{\parallel} Q_2^{\parallel}$,
where
$t_{\perp} = 2 t_1^{\perp}({\hat{\bi x}}) R_1^{\perp}$, and where
$t_{\perp}^{\prime} = 4 t_2^{\perp}({\hat{\bi x}}+{\hat{\bi y}}) R_2^{\perp}/i$
are real hopping amplitudes.
Above, $\phi$ denotes the angle that ${\bi k}$ makes with $x$ axis.
Henceforth, we shall assume hole bands: $t_{\parallel} < 0$.
The constant energy contours
$\epsilon_F = \varepsilon_{\parallel}({\bi k}) \pm |\varepsilon_{\perp} ({\bi k}) |$
then yield Fermi surfaces
\begin{equation}
k_{F \pm}^2 (\phi) a^2 = \epsilon_F^{\prime} /
(-t_{\parallel}  \pm 
[t_{\perp}^2 \cos^2 (2\phi) + t_{\perp}^{\prime 2} \sin^2 (2\phi)]^{1/2})
\label{FSS}
\end{equation}
at low doping $x\ll 1$, where
$\epsilon_F^{\prime} = \epsilon_F-\varepsilon_{\parallel}(0)$.
These are shown in Fig. \ref{FS}.
The average of the volume of phase space contained by the inner ($+$) Fermi surface
and the outer ($-$) Fermi surface
gives the hole concentration per orbital:
$x = (2\pi)^{-2}\int_0^{2\pi} d\phi \sum_{\pm} \int_0^{k_{F \pm}(\phi)} dk\, k\, a^2/2$.
The resulting angle integral is obtained by
applying the residue theorem after  making
the change of variable $z = e^{4i\phi}$,
which yields the relationship
\begin{equation}
4\pi x = \epsilon_F^{\prime} (-t_{\parallel}) / (t_{\parallel}^2 - t_{\perp}^2)^{1/2}
(t_{\parallel}^2 - t_{\perp}^{\prime 2})^{1/2}
\label{e_f_vs_x}
\end{equation}
between the Fermi energy and the concentration of mobile holes.
Next, the intra-orbital hopping fields for slave fermions are given by
$P_n^{\parallel} = 
(2\pi)^{-2}\int_0^{2\pi} d\phi \sum_{\pm} 
\int_0^{k_{F \pm}(\phi)} dk\, k\, (1 - {n\over 4} k^2 a^2)\, a^2/4$ at $x\ll 1$.
Performing the angle integral after the previous change of variable yields the result
\begin{equation}
P_n^{\parallel} = {1\over 2} x - {n \pi\over 4} x^2
{t_{\parallel}^2 - t_{\perp}^2 t_{\perp}^{\prime 2}/t_{\parallel}^2\over{(t_{\parallel}^2 - t_{\perp}^2)^{1/2}
(t_{\parallel}^2 - t_{\perp}^{\prime 2})^{1/2}}}
\label{P_para_2}
\end{equation}
for these amplitudes.
Last, the mean fields for inter-orbital hopping by slave fermions are given by
$P_1^{\perp}(\hat{\bi x}) =
(2\pi)^{-2}\int_0^{2\pi} d\phi \sum_{\pm}
\int_0^{k_{F \pm}(\phi)} dk\, k^3\, (\mp\cos \, 2 \phi) e^{2i\delta_f({\bi k})}   \, a^4/16$
and by
$P_2^{\perp}(\hat{\bi x} + \hat{\bi y}) =
(2\pi)^{-2}\int_0^{2\pi} d\phi \sum_{\pm}
\int_0^{k_{F \pm}(\phi)} dk\, k^3\, (\mp\sin \, 2 \phi) e^{2i\delta_f({\bi k})}   \, a^4/8$
at low hole concentrations,
where
\begin{equation}
e^{2i\delta_f({\bi k})} = - {t_{\perp} \cos (2\phi) + i t_{\perp}^{\prime} \sin (2\phi)\over{
[t_{\perp}^2 \cos^2 (2\phi) + t_{\perp}^{\prime 2} \sin^2 (2\phi)]^{1/2}}}
\label{phase_factor_at_0}
\end{equation}
is the fermion phase factor at ${\bi k}$ near zero.  
Employing again the change of variable $z = e^{4i\phi}$ in the resulting angle integrals then yields
the following expressions for these amplitudes:
\begin{eqnarray}
 P_1^{\perp}(\hat{\bi x})  = 
{\pi\over 4} x^2 {t_{\perp}\over{t_{\parallel}}}
\Biggl({t_{\parallel}^2- t_{\perp}^{\prime 2}\over{t_{\parallel}^2- t_{\perp}^2}}\Biggr)^{1/2}, 
\label{P_perp_1} \\
 P_2^{\perp}(\hat{\bi x} + \hat{\bi y})  = 
i{\pi\over 2} x^2 {t_{\perp}^{\prime}\over{t_{\parallel}}}
\Biggl({t_{\parallel}^2- t_{\perp}^2\over{t_{\parallel}^2- t_{\perp}^{\prime 2}}}\Biggr)^{1/2}.
\label{P_perp_2}
\end{eqnarray}
Substitution into the expression
for the inter-orbital matrix element 
experienced by Schwinger bosons (\ref{Omega_perp_prime_1}) 
reduces it to
\begin{eqnarray}
\Omega_{\perp}^{\prime}({\bi k}) =
\pi x^2 R_1^{\perp} {|2 t_1^{\perp}|^2\over{t_{\parallel}}}
\Biggl({t_{\parallel}^2- t_{\perp}^{\prime 2}\over{t_{\parallel}^2- t_{\perp}^2}}\Biggr)^{1/2}
\gamma_1({\bi k})
+ \pi x^2 R_2^{\perp} {|4 t_2^{\perp}|^2\over{t_{\parallel}}}
\Biggl({t_{\parallel}^2- t_{\perp}^2\over{t_{\parallel}^2- t_{\perp}^{\prime 2}}}\Biggr)^{1/2}
\gamma_2({\bi k}). \nonumber \\
\label{Omega_perp_prime_2}
\end{eqnarray}
Observe that the phase factor associated with Schwinger bosons is then
$e^{2i\delta_b ({\bi k})} = {\rm sgn}\, t_{\parallel} = -1$
near ${\bi k} = 0$.  The principal mean-field equation (\ref{s_0_mf_eq}) implies
ideal Bose-Einstein condensation (BEC) of the Schwinger bosons into the bottom of
the spectrum (\ref{w_b}) at $3$-momentum
$(k_0, {\bi k}) = (\pi, 0)$ in the zero-temperature limit.
Comparison of (\ref{s_0_mf_eq})
with the mean-field equations (\ref{Q_para_mf_eq}) and (\ref{Q_perp_mf_eq})
yields the mean-field values
$Q_n^{\parallel} = s_0$ and 
$Q_n^{\perp} = s_1$, 
in the large-$s_0$ limit,
where $s_1 =  s_0 \tanh \, 2 \theta = s_0 [\Omega_{\perp}(0) / \Omega_{\parallel}(0)]$. 
Ideal BEC necessarily requires $\omega _b (\pi,0) = 0$. 
This yields ultimately that 
$\tanh \, 2 \theta  = (1+[\Omega_{\perp}^{\prime}(0)/\Omega_{\perp}(0)]^2)^{-1/2}$.  
Inter-orbital hopping therefore diminishes the antiferromagnetic order 
in the hidden half-metal state: $s_1 < s_0$.
Notice, however, that the correction is small and of order $x^4$.
Last, ideal BEC implies a unique mean-field amplitude
$R_n^{\perp} = s_2$ by meanfield equations (\ref{R_perp_mf_eq}),
where $s_2 = s_0 \, {\rm sech} \, 2 \theta$,
with ${\rm sech} \, 2 \theta = 
[|\Omega_{\perp}^{\prime}(0)|/\Omega_{\perp}(0)]/(1+[\Omega_{\perp}^{\prime}(0)/\Omega_{\perp}(0)]^2)^{1/2}$.
Observe now that the latter ratio of frequencies takes the form
$|\Omega_{\perp}^{\prime}(0)|/\Omega_{\perp}(0) = A_0 (s_2 / s_0 s_1)$ as $x\rightarrow 0$, 
in which case $s_1\rightarrow s_0$.
Here, $A_0$ is a positive constant that is small compared to unity at low hole concentration.
Substituting this form into the previous definition of $s_2$ yields only the trivial solution $s_2 = 0$.
We therefore conclude that inter-orbital hopping is dynamically suppressed in the present
Schwinger-boson-slave-fermion mean-field theory for the hidden half-metal state
at large electron spin $s_0$ and low hole concentration.
In particular, 
Eqs. (\ref{P_perp_1}) and (\ref{P_perp_2}) imply that all of
the inter-orbital mean fields, $P_n^{\perp}$ and $R_n^{\perp}$, are null. 

Dynamical suppression of inter-orbital hopping at large-$s_0$ implies a null
inter-orbital hopping matrix element for Schwinger bosons by Eq. (\ref{Omega_perp_prime_2}): 
$\Omega_{\perp}^{\prime} ({\bi k}) = 0$.
Inspection of the spectrum for Schwinger bosons (\ref{w_b})
then yields the dispersion relation
\begin{equation}
\omega_b(k_0, {\bi k}) = v_0 |{\bi k}|
\quad {\rm near}\quad  {\bi k} = 0
\label{k_0_dispersion}
\end{equation}
at the zero-temperature limit, where
\begin{equation}
v_0 = 2 s_0 a (1-x)^2 ([J_1^{\perp} - {\bar J}_1^{\parallel}(x) + 2 J_2^{\perp} - 2 {\bar J}_2^{\parallel}(x)]
\cdot [{1\over 2} J_0 + 2 J_1^{\perp} + 2 J_2^{\perp}])^{1/2}
\label{v0}
\end{equation}
is the velocity of hidden spinwaves,
with  ${\bar J}_n^{\parallel}(x) = J_n^{\parallel} +  (1-x)^{-2} s_0^{-1} t_n^{\parallel} x$
for $n=1, 2$.
At ${\bi k}$ near cSDW wavenumbers $(\pi/a)\hat{\bi x}$ and $(\pi/a)\hat{\bi y}$,
the spectrum for Schwinger bosons (\ref{w_b}) disperses anisotropically as 
\begin{equation}
\omega_b(k_0, {\bi k}) =
[\Delta_{cSDW}^2 + v_{l}^2 (k_{l} - \pi / a)^2 + v_{t}^2 k_{t}^2]^{1/2}
\label{csdw_dispersion}
\end{equation}
in the zero-temperature limit, with a spin gap
\begin{equation}
\Delta_{cSDW} = (1-x)^2 (2 s_0) [(4 J_2^{\perp} - J_{0c}) (J_0 - J_{0c})]^{1/2}
\label{delta_csdw}
\end{equation}
that vanishes at a critical Hund's Rule coupling 
%
\begin{equation}
- J_{0c} = 2 (J_1^{\perp} - J_1^{\parallel}) - 4 J_2^{\parallel} -
  (1-x)^{-2} s_0^{-1} 2  t_{\parallel} x .
\label{J_0c}
\end{equation}
Above, $k_l$ and $k_t$ denote the longitudinal and transverse components
of ${\bi k}$ with respect to the cSDW momentum.
At criticality, $\Delta_{cSDW} = 0$, 
the longitudinal spinwave velocity $v_l$ coincides with the hidden spinwave velocity $v_0$,
while the anisotropy parameter
is given by
\begin{equation}
{v_l\over{v_t}} = \Biggl[{2 {\bar J}_2^{\parallel}(x) + 2 J_2^{\perp} + {\bar J}_1^{\parallel}(x) + J_1^{\perp}\over
{2 {\bar J}_2^{\parallel} (x) + 2 J_2^{\perp} - {\bar J}_1^{\parallel}(x) - J_1^{\perp}}}\Biggr]^{1/2},
\label{ani_csdw}
\end{equation}
which is greater than unity.
The hidden half metal is stable at weak to moderate Hund's Rule coupling $-J_0 < -J_{0c}$.
Equation (\ref{J_0c}) therefore implies that
intra-orbital hole hopping
stabilizes the hidden half-metal state for hole bands, $t_{\parallel} < 0$.
(See Fig. \ref{phase_diagram}.)
We conclude that the hidden half metal is robust with 
respect to the presence of inter-orbital hopping
within the present Schwinger-boson-slave-fermion mean-field theory.

Finally,
returning to the generic unoptimized mean-field theory, 
the inversion of the similarity transform (\ref{rotate_f}) yields slave-fermion states
that are annihilated by
\begin{equation}
f(k_0,{\bi k}) = 
{e^{+i\delta_f({\bi k})}\over{2^{1/2}}}
f_{d-}({\bi k}) +
e^{ik_0} {e^{-i\delta_f({\bi k})}\over{2^{1/2}}}
f_{d+}({\bi k}) .
\end{equation}
Their corresponding Wannier wave functions
then depend on the azimuthal angle $\phi^{\prime}$ about each iron atom  as
$\cos [\phi^{\prime}-\delta_f({\bi k})]$
 in the even channel, $k_0 = 0$, and as
$\sin [\phi^{\prime}-\delta_f({\bi k})]$
in the odd channel, $k_0 = \pi$.
Figure \ref{FS} depicts slave-fermion Fermi surfaces (\ref{FSS}) at low doping $x = 0.01$. 
Dynamical suppression of inter-orbital hopping implies that the matrix element for
inter-orbital hopping of slave fermions (\ref{epsilon_perp}) is null, however.
The inner and outer Fermi surface hole pockets depicted by Fig. \ref{FS} must therefore
collapse into two degenerate circular Fermi surface hole pockets 
at low doping, $x\ll 1$, at large spin $s_0$.
We believe that this effect is due to the 
inability of the two-orbital $t$-$J$ model (\ref{tJ}) to
``erase'' strings of overturned spins that result from 
inter-orbital hole hopping in the classical limit, at large spin $s_0$\cite{S_S_88}\cite{Trugman_88}.
Related behavior is predicted theoretically for
a mobile hole in the N\' eel state over the one-orbital square lattice, where nearest-neighbor
hopping ($t_1$) is dynamically suppressed at low energy, 
leaving effective next-nearest-neighbor
hopping ($t_2$) within the 
same antiferromagnetic sublattice\cite{Auerbach_bk}\cite{Auerbach_93}\cite{Kane_90}.
As we will see in section 4, 
this limiting result persists
for true electron spin $s_0 = 1/2$ in a special case.

\begin{figure}
\includegraphics[scale=1.2]{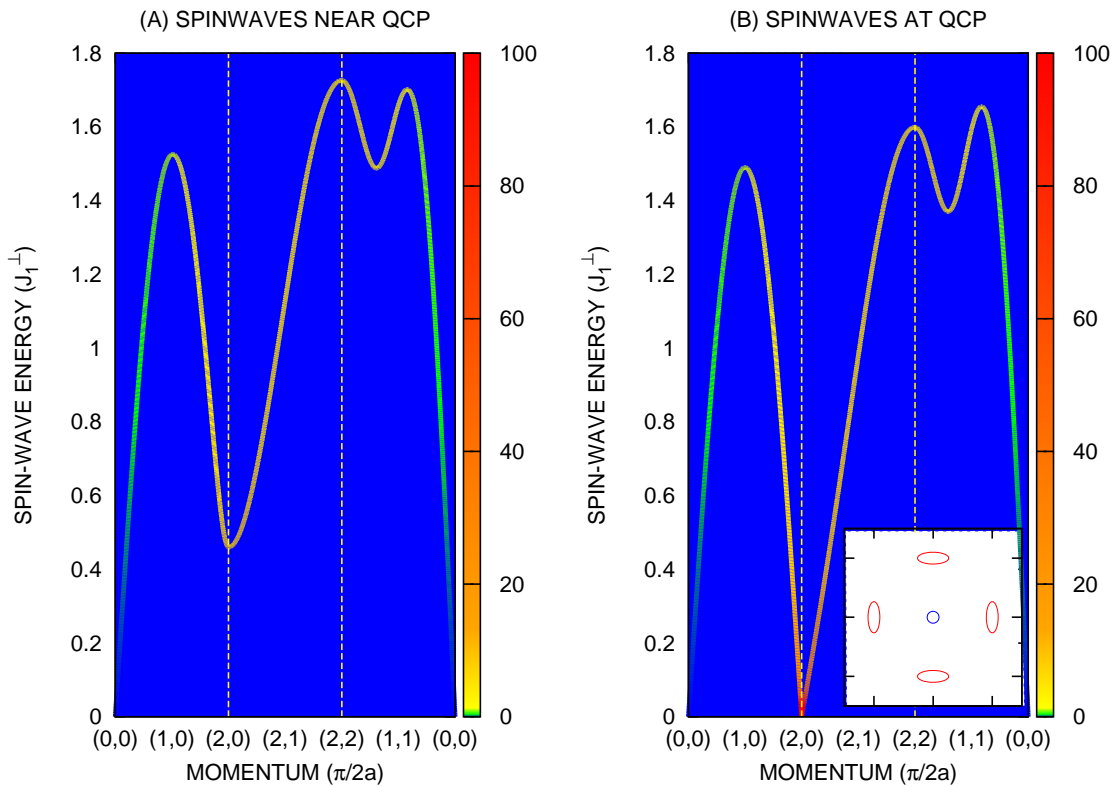}
\caption{The mean-field result for the
imaginary part of the dynamical spin response function
in the zero-temperature-large-$s_0$ limit,
Eqs. (\ref{chi_perp}) and (\ref{spctrl_wght}),
is evaluated  with the following set of parameters:
$J_1^{\parallel} = 0$, $J_1^{\perp} > 0$,
$J_2^{\parallel} = 0.3\, J_1^{\perp} = J_2^{\perp}$,
$t_1^{\parallel} = -5 J_1^{\perp}$, $t_2^{\parallel} = 0$,
$x = 0.01$, and $s_0 = 1/2$.
Dynamical suppression of inter-orbital hopping is presumed.
The Hund's Rule coupling is set to $-J_0 = -J_{0c} - 0.1 J_1^{\perp}$ just off the QCP in panel (A),
while it is set to the critical value $-J_{0c}$ in panel (B).
Low-energy contours are displayed in the inset.}
\label{spin_x}
\end{figure}

\subsection{Spinwaves}
We shall now determine the spin excitations of 
the hidden half metal by directly computing the dynamical
spin-spin correlation function from the Schwinger-boson-slave-fermion mean-field theory.  First,
we identify the true $(k_0 = 0)$ versus the hidden $(k_0 = \pi)$ spin at each iron atom $i$:
${\bi S}_i (k_0) = {\bi S}_{i,d-} + e^{i k_0}{\bi S}_{i,d+}$.
Consider then the transverse dynamical spin-spin correlation function
$\langle {\bi S} (k_0) \cdot {\bi S}^{\prime} (k_0) \rangle |^{\perp}
= {1\over 2} \langle S^{+} (k_0) S^{\prime -} (k_0)
+ S^{-} (k_0) S^{\prime +} (k_0) \rangle$ at space-time points
$({\bi r},t)$ and $({\bi r}^{\prime},t^{\prime})$.
Averaging the slave-fermion dynamics over the bulk, we obtain the form 
\begin{eqnarray}
\langle {\bi S} (k_0) \cdot {\bi S}^{\prime} (k_0) \rangle |^{\perp}_{{\bi k},\omega} =
(1-x)^2 {\hbar^2\over 2}
& \sum_{\alpha = 0,1}\sum_{\beta = 0,1} e^{ik_0(\alpha-\beta)} \cdot \nonumber \\
& \cdot [G_{\alpha,\beta}^{(b)} * G_{\alpha,\beta}^{(b)*} +
F_{\alpha,\beta}^{(b)} * F_{\alpha,\beta}^{(b)*}]|_{{\bi k},\omega},
\end{eqnarray}
for its Fourier transform, where
$i G_{\alpha,\beta}^{(b)} ({\bi r}_i,t;{\bi r}_j,t^{\prime}) = 
\langle b_{i,\alpha,s}(t) b_{j,\beta,s}^{\dagger}(t^{\prime})\rangle$
and
$i F_{\alpha,\beta}^{(b)} ({\bi r}_i,t;{\bi r}_j,t^{\prime}) = 
\langle b_{i,\alpha,s}(t) b_{j,\beta,\bar s}(t^{\prime})\rangle$
are the regular and the anomalous Greens functions for the Schwinger bosons,
and where the notation  $f*g$ denotes a convolution in frequency and momentum.
Above, we use the index $0$ and $1$ for the orbitals $d-$ and $d+$.
Also, henceforth in this subsection, we shall identify $k_0 = 0$ and $\pi$
with the respective labels $+$ and $-$.
After substitution of the Bogoliubov transformation (\ref{Bgv_trns_a},$b$)
and of the similarity transform (\ref{sim_trans_b}),
a standard summation of the Matsubara frequencies in the convolution yields the following
Auerbach-Arovas expression for
the dynamical spin correlator at $T>0$ \cite{Auerbach_Arovas_88}:
\begin{eqnarray}
i\langle {\bi S}(\pm) \cdot {\bi S}^{\prime} (\pm) \rangle |^{\perp}_{{\bi k},\omega} =
{\hbar^2\over 4}  {(1-x)^2\over{\cal N}} & \sum_{q_0 = +, -} \sum_{\bi q} \Biggl\{
 (1 + \cosh \, [2\theta({\bi q}) \mp 2\theta({\bi q}-{\bi k})])\cdot \nonumber \\
&\cdot\sum_{\nu = -\omega,+\omega} {n_B[\omega_b(q_0,{\bi q})]-n_B[\omega_b(q_0 \pm,{\bi q}-{\bi k})]\over{\nu - \omega_b(q_0,{\bi q}) + \omega_b(q_0 \pm,{\bi q}-{\bi k})}}
+ \nonumber \\
&+(1 - [\cosh \, 2\theta({\bi q})] [\cosh \, 2\theta({\bi q}-{\bi k})])\cdot \nonumber \\
&\cdot\sum_{\nu = -\omega,+\omega} {n_B[\omega_b(q_0,{\bi q})]+n_B[\omega_b(q_0 \pm,{\bi q}-{\bi k})]\over{\nu - \omega_b(q_0,{\bi q}) - \omega_b(q_0 \pm, {\bi q}-{\bi k})}}
+ \nonumber \\
& \pm [\sinh \, 2\theta({\bi q})] [\sinh \, 2\theta({\bi q}-{\bi k})]\cdot \nonumber \\
&\cdot\sum_{\nu = -\omega,+\omega} {n_B[\omega_b(q_0,{\bi q})]+n_B[\omega_b(q_0 \mp,{\bi q}-{\bi k})]\over{\nu - \omega_b(q_0,{\bi q}) - \omega_b(q_0 \mp, {\bi q}-{\bi k})}} \Biggr\}. \nonumber \\
\label{AA}
\end{eqnarray}
%
When computing $q_0 \pm$ above, we use the multiplication table $\pm \pm = +$ and $\mp\pm = -$.
The property
$\omega_b(q_0, {\bi q}) = \omega_b(q_0, -{\bi q})$
displayed by the spectrum for Schwinger bosons (\ref{w_b})
was exploited to obtain the reduced expression (\ref{AA}).
Expression (\ref{AA}) can be used to show that
$(2\pi)^{-1}\int_{-\infty}^{+\infty} d\omega
\langle {\bi S}(+) \cdot {\bi S}^{\prime}(+) \rangle |^{\perp}_{{\bi k},\omega} = 0$ 
at ${\bi k}= 0$.
This  is consistent
with a spin-singlet hidden-order antiferromagnetic state
inside the subspace where $\sum_i\sum_{\alpha} S_{z,i,\alpha} = 0$:
$\langle |\sum_i\sum_{\alpha} {\bi S}_{i,\alpha}|^2\rangle = 0$.

At large electron spin $s_0$,
the true self-consistent mean-field theory dynamically suppresses inter-orbital hopping, however.
This leads to Schwinger bosons with degenerate spectra:
$\omega_b(q_0, {\bi k}) = \omega_b^{(0)} ({\bi k})$ for $q_0 = +,-$.
At zero temperature,
expression (\ref{AA}) in conjunction with ideal BEC then yields the result\cite{jpr11}
\begin{equation}
i \langle {\bi S}(\pm) \cdot {\bi S}^{\prime}(\pm) \rangle |^{\perp}_{{\bi k},\omega}
= \pi^{-1} A(\pm,{\bi k})
([\omega_b^{(0)} ({\bi k}) - \omega ]^{-1} + [\omega_b^{(0)}({\bi k}) + \omega ]^{-1}),
\label{chi_perp}
\end{equation}
where the poles in frequency have spectral weight 
\begin{equation}
A(\pm,{\bi k}) = \pi (1-x)^2 s_0\hbar^2
[\Omega_{\mp}({\bi k})/\Omega_{\pm}({\bi k})]^{1/2}.
\label{spctrl_wght}
\end{equation}
Here, $\Omega_{\pm} = \Omega_{\parallel}\pm\Omega_{\perp}$.
The above dynamical spin correlator coincides with the transverse spin
susceptibility, $\chi_{\perp}(k,\omega)$, in the present zero-temperature limit
by the fluctuation-dissipation theorem.
Notice that 
the spectral weights (\ref{spctrl_wght}) above
diverge at $\Omega_{\pm}({\bi k}) = 0$.
This implies the  existence of hidden $(-)$ spin waves 
near zero 2D momentum
with a spectrum (\ref{k_0_dispersion}),
as well as true $(+)$  spin waves near cSDW momenta
with a spectrum (\ref{csdw_dispersion}) \cite{jpr11}.
We conclude that the hidden half-metal state 
shows strict long-range antiferromagnetic
order across the $d+$ and $d-$ orbitals of the iron atoms
at zero temperature.

The true spin excitations $(+)$ predicted by 
the dynamical spin susceptibility (\ref{chi_perp})
near the quantum-critical point are shown graphically 
by Fig. \ref{spin_x} for a set of parameters 
that is applicable to iron-pnictide high-temperature superconductors.
Notice the spin gap (\ref{delta_csdw}) that exists at cSDW wavenumbers
$(\pi / a){\hat{\bi x}}$ and  $(\pi / a){\hat{\bi y}}$.
It collapses to zero at a critical Hund's Rule coupling (\ref{J_0c})
of moderate strength.
This suggests that the hidden half-metal state gives way to a cSDW metal phase\cite{ran_09}
that shows long-range cSDW correlations
at the QCP, which separates the two phases.

\subsection{Fermi Surfaces} 
We shall now obtain the electronic structure of the
hidden half-metal state by computing the one-electron
propagator directly from
the Schwinger-boson-slave-fermion
mean-field theory for the two-orbital $t$-$J$ model.
It is given by the convolution of the propagator for
Schwinger bosons with the propagator for slave fermions:
$i G({\bi k},\omega) = \sum_{\alpha=d-, d+} G_{\alpha,\alpha}^{(b)} 
* G_{\alpha,\alpha}^{(f)*} |_{{\bi k},\omega}$,
where 
$i G_{\alpha,\beta}^{(f)}({\bi r}_i,t;{\bi r}_j,t^{\prime})
=\langle f_{i,\alpha}(t) f_{j,\beta}^{\dagger}(t^{\prime})\rangle$.
A standard summation of Matsubara frequencies  yields the expression\cite{jpr11}
\begin{eqnarray}
G({\bi k},\omega) = \sum_{k_0=0,\pi}{1\over{\cal N}}\sum_q 
&\Biggl( 
[\cosh \, \theta({\bi q})]^2
{{n_B[\omega_b (q)] + n_F[\varepsilon_f (q-k)]}\over{\omega - \omega_b(q) + \varepsilon_f(q-k)}} +
\nonumber \\
& +[\sinh \, \theta({\bi q})]^2
{{n_B[\omega_b (q)] + {\bar n}_F[\varepsilon_f(q-k)]}\over{\omega + \omega_b(q) + \varepsilon_f(q-k)}}\Biggr) .
\label{G}
\end{eqnarray}
Above, 
${\bar n}_F(\varepsilon) = ({\rm exp}[(\mu-\varepsilon)/k_B T] + 1)^{-1}$.
All of the Schwinger bosons condense
into the lowest-energy state
at  $3$-momentum $(q_0, {\bi q}) = (\pi, 0)$
as $T\rightarrow 0$
by the principal meanfield equation (\ref{s_0_mf_eq}).
This results in the following
coherent contribution to the electronic spectral function
at zero temperature and at large $s_0$:
\begin{equation}
{\rm Im}\, G_{\rm coh}({\bi k},\omega) = s_0
{\pi} \sum_{k_0=0,\pi}\delta[\omega + \varepsilon_f(k_0,{\bi k})].
\label{im_g_coh}
\end{equation}
In the case of the unoptimized generic mean-field theory,
it reveals inner and outer hole Fermi surfaces centered at zero 2D  momentum
that are depicted by Fig. \ref{FS}.
These collapse into doubly-degenerate circular hole pockets
in the self-consistent mean-field theory at large $s_0$,
where inter-orbital hopping is dynamically suppressed.


The contribution due to the Fermi-Dirac terms in the above expression (\ref{G})
represent incoherent excitations in the electronic structure.
At energies $\omega$ below the electronic Fermi level,
the pole in the second term of expression (\ref{G}) represents 
the combination of a hole excitation, $\varepsilon_f > \mu$,
with a spinwave, $\omega_b > 0$.
This composite excitation
therefore shows a gap (\ref{delta_csdw}) 
about cSDW momenta, $\Delta_{cSDW}$.
Notice that copies of the previous inner and outer Fermi surfaces 
now centered at cSDW wavenumbers
$(\pi / a){\hat{\bi x}}$ and $(\pi / a){\hat{\bi y}}$
exist at the QCP because of the collapse of the spin gap there: $\Delta_{cSDW}\rightarrow 0$.
(See Fig. \ref{spin_x}.) This is a spin-density wave nesting mechanism in reverse,
where low-energy spinwaves centered at cSDW momenta produce nested Fermi surfaces.
The incoherent contribution due to the first term in expression (\ref{G})
above vanishes in the zero-temperature limit at energies $\omega$ below the
electronic Fermi level.
Detailed evaluations of ${\rm Im}\, G_{\rm inc}({\bi k},\omega)$ 
from the Fermi-Dirac terms in expression (\ref{G})
find electron-type dispersion for the emergent band at momenta 
that lie inside of the nested Fermi surface pocket\cite{jpr11}.
They also find hole-type dispersion of the emergent band at momenta that lie outside of the
nested Fermi surface\cite{jpr11}, however.  We believe that the latter 
is an artifact of the present mean-field approximation.  
Indeed, strong arguments for electron-type emergent bands will be 
presented at the end of section 5.


To conclude, Schwinger-boson-slave-fermion meanfield theory
of the two-orbital $t$-$J$ model (\ref{tJ})
at a quantum-critical point 
predicts nested Fermi surfaces centered
at zero 2D momentum and at cSDW momenta that are circular in shape.
The Fermi surface pockets centered at cSDW momenta
have purely $3d_{xz}$ and $3d_{yz}$ orbital character by construction,
which is consistent with a recent ARPES study of the electronic structure in
optimally doped Ba(Fe$_{1-x}$Co$_x$)$_2$As$_2$ \cite{zhang11}.

\section{Exact Diagonalization} 
The previous Schwinger-boson-slave-fermion meanfield theory for the two-orbital $t$-$J$ model (\ref{tJ})
predicts that electronic structure centered at cSDW momenta
emerges at a quantum critical point that separates a cSDW metal
from a  hidden half-metal state.
This result was obtained in the limit of large electron spin $s_0$.
Below, we will see that the effect persists
for true electron spin, $s_0 = 1/2$.

We have obtained  the exact low-energy spectrum of the two-orbital $t$-$J$ model (\ref{tJ})
for  $4\times 4\times 2$ local spin-$1/2$ moments plus one mobile hole
by computer calculation.
The Schwinger-boson-slave-fermion representation (\ref{constraint})
of the correlated electron is applied exactly.
In particular,
a quantum state is specified by the combination of 
an arbitrary spin background over all of the
$4\times 4\times 2$ sites,
confined to the subspace with total spin $S_z = 0$,
and a hole location at one of the down-spin sites in the given spin background.
The Heisenberg-exchange and Hund-exchange terms in the $t$-$J$ model Hamiltonian (\ref{tJ})
reduce to permutations of the spin backgrounds, and these are  stored in memory.
The matrix elements for  correlated hopping terms in (\ref{tJ}) are  computed directly
at each application of the Hamiltonian operator,
on the other hand.
Periodic boundary conditions are imposed,
and   translation and reflection symmetries on the $4\times 4\times 2$
lattice are exploited in order to reduce the dimension of the Hilbert space.
Global swap of the orbitals, $P_{d\bar d}$, is included in the list of symmetry
operations at the extreme where inter-orbital next-nearest-neighbor hopping
is suppressed: $t_2^{\perp} = 0$.
For example,
exploiting reflections about both principal axes of the $4\times 4$ square lattice of iron atoms
plus $P_{d\bar d}$
brings the dimension of the Hilbert space down to 75,624,211 states
in such case at  zero-momentum and  even reflection parities.
The low-energy states of the resulting block-diagonal Hamiltonian are 
obtained by employing the Lanczos technique\cite{lanczos}.  
We use the ARPACK subroutine library for this purpose\cite{arpack},
in which case the Hamiltonian operation 
$|\psi^{\prime}\rangle = H |\psi\rangle$
is  accelerated by exploiting parallel threads with OpenMP directives.

\begin{figure}
\includegraphics{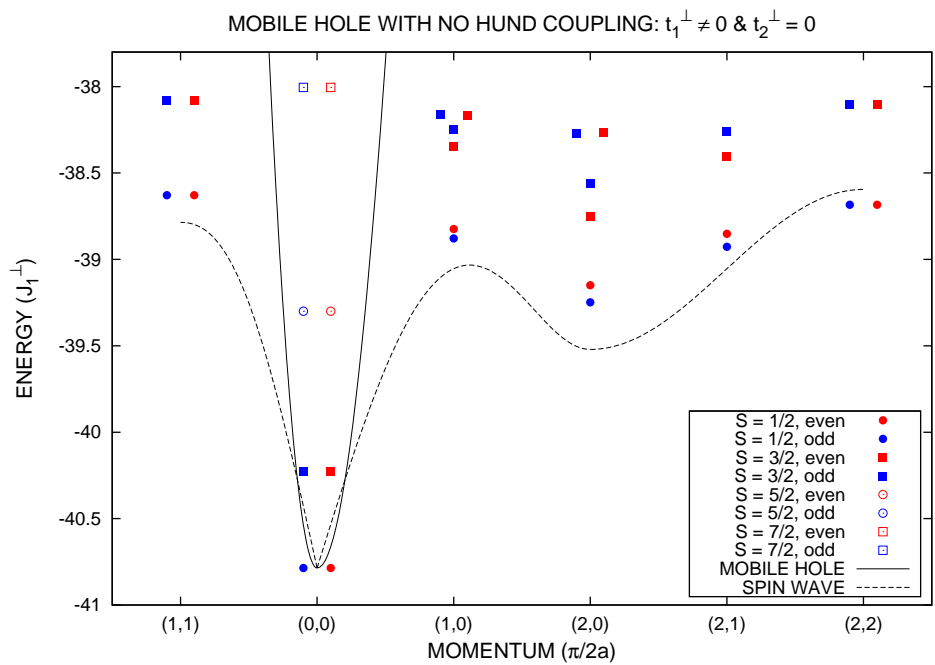}
\caption{Shown is the low-energy spectrum of the two-orbital $t$-$J$ model,
Eq. (\ref{tJ}),  over a
$4\times 4\times 2$  lattice with one hole,
in the absence of Hund's Rule: $J_0 = 0$. 
The remaining parameters are 
$J_1^{\parallel} = 0$, 
$J_1^{\perp} > 0$,
$J_2^{\parallel} = 0.3 J_1^{\perp} = J_2^{\perp}$,
$t_1^{\parallel} = - 5 J_1^{\perp}$,
$t_1^{\perp} ({\hat{\bi x}}) = - 2 J_1^{\perp}$, 
$t_1^{\perp} ({\hat{\bi y}}) = + 2 J_1^{\perp}$, and
$t_2^{\parallel} = 0 = t_2^{\perp}$.
Red and blue states are respectively even and odd under $P_{d\bar d}$.
Henceforth, some points on the spectrum
are artificially moved slightly off their quantized values along the
momentum axis for the sake of clarity.
A comparison with the hole spectrum, 
$\varepsilon_f(k) = \varepsilon_{\parallel}({\bi k})$, 
and with the spin-wave spectrum,
$\omega_b(k) = \omega_b^{(0)}({\bi k})$, 
at large $s_0$ and  $x=0$ is also shown.}
\label{spctrm_wc_a}
\end{figure}

\begin{figure}
\includegraphics{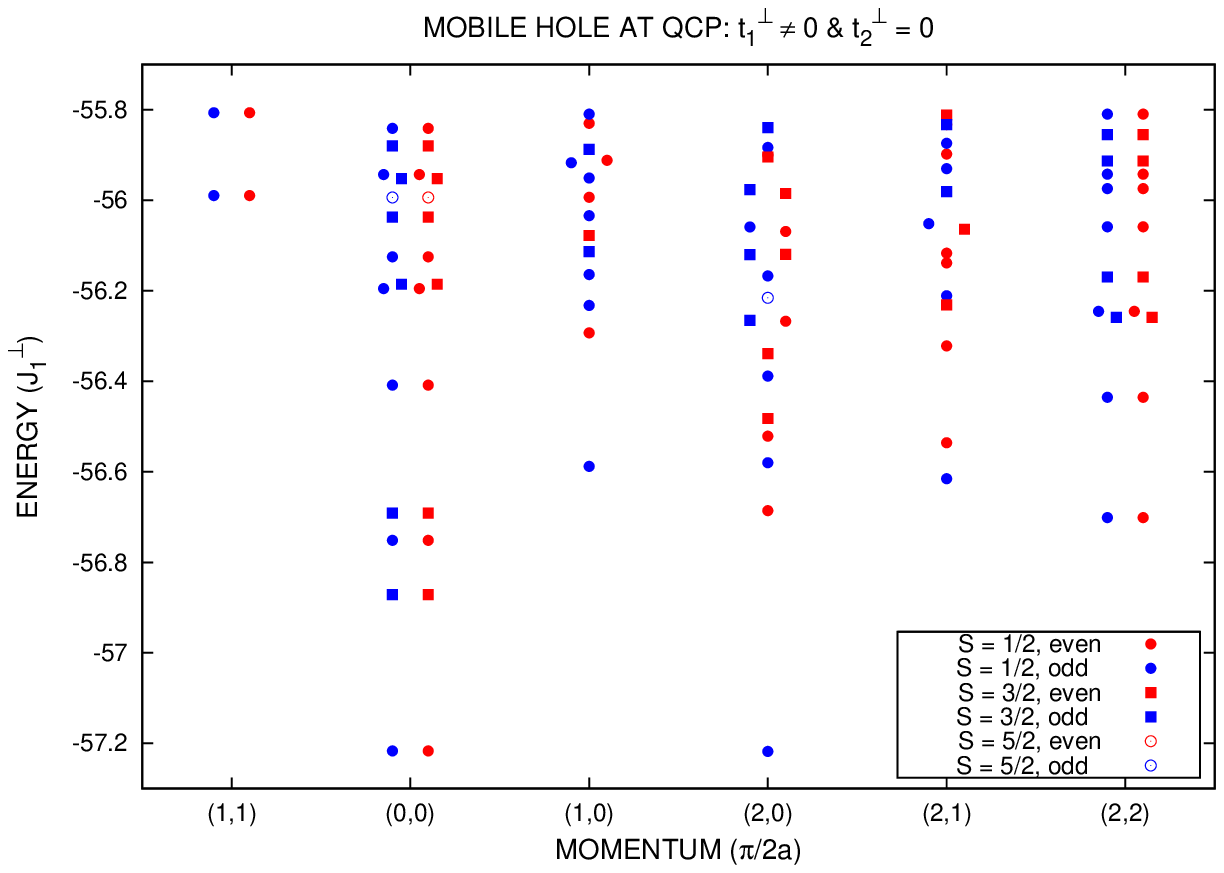}
\caption{The low-energy spectrum of the two-orbital $t$-$J$ model with one hole roaming
over a $4\times 4\times 2$  lattice is displayed at the QCP that separates the cSDW from the
hidden half-metal state: $J_{0c} = -1.733 J_1^{\perp}$. The remaining set of parameters are listed
in the caption to Fig. \ref{spctrm_wc_a}.
Red and blue states are respectively even and odd under $P_{d\bar d}$.}
\label{spctrm_cp_a}
\end{figure}

Figures \ref{spctrm_wc_a}, \ref{spctrm_cp_a} and \ref{spctrm_sc_a} display the evolution of the
low-energy spectrum of one mobile hole in a $4\times 4$ lattice of spin-$1$ iron atoms 
with the strength of the Hund's Rule coupling, $-J_0$.
We chose the following set of Heisenberg exchange coupling constants
and hopping matrix elements:
$J_1^{\parallel} = 0$, $J_1^{\perp} > 0$, $J_2^{\parallel} = 0.3 J_1^{\perp} = J_2^{\perp}$,
$t_1^{\parallel} = - 5 J_1^{\perp}$, 
$t_1^{\perp}({\hat{\bi x}}) = -2 J_1^{\perp}$,  $t_1^{\perp}({\hat{\bi y}}) = +2 J_1^{\perp}$,
and $t_2^{\parallel} = 0 = t_2^{\perp}$.
Recall that global swap of the orbitals, $P_{d\bar d}$, is then an exact symmetry. 
(See the end of section 2.)
Also, the mean-field result (\ref{J_0c}) then predicts a quantum critical point at
$J_0 = - 0.8 J_1^{\perp}$ in the thermodynamic limit $(x = 0)$.
It separates a hidden half-metal state at weak Hund's Rule coupling
from a cSDW at strong Hund's Rule coupling.
Figure \ref{spctrm_wc_a} displays the exact spectrum in the absence of Hund's Rule.
Red and blue points denote states that are respectively even and odd under $P_{d\bar d}$.
The spectrum is consistent with a hidden half metal.  
In particular, the lowest-energy states at fixed momentum carry spin-$1/2$
and they are nearly doubly degenerate.
Furthermore,
the pairs of spin-$1/2$ states in Fig. \ref{spctrm_wc_a} are
close in energy to
the next-excited pairs of states at fixed momentum, which carry spin-$3/2$.
By the summation of angular momentum,
these two pairs of states can be understood
as  the combination of a well-defined spin-$1/2$ hole excitation
and  a well-defined spin-$1$ excitation that interact weakly.
The pole in the second term of the one-electron propagator (\ref{G})
obtained by the previous Schwinger-boson-slave-fermion meanfield theory
for the hidden half-metal state
suggests that the energy dispersion relation of one hole of momentum ${\bi q}$ is equal to
$\varepsilon_f(0) + \omega_b^{(0)}({\bi q})$ at $v_0/a \ll -t_{\parallel}$.
In other words, the momentum of one hole is carried entirely by the spinwave
that accompanies it.
Figure \ref{spctrm_wc_a} shows that the dispersion of the pairs of spin-$1/2$ and spin-$3/2$
low-energy  states  follows rather closely the trend set by
the spin-wave dispersion $\omega_b^{(0)}({\bi q})$. (See also Fig. \ref{spctrm_wc_b}.)
In particular, the mean-field result sets a lower bound for the exact dispersion
of one hole in the two-orbital $t$-$J$ model (\ref{tJ}).
This is very likely due to the larger Hilbert space of the
Schwinger-boson-slave-fermion mean-field theory, which only enforces the
constraint against double occupancy (\ref{constraint}) on average over the bulk.

%
\begin{table}
\caption{\label{table1} 
Listed are groundstate expectation values of physical observables
for one hole hopping over a $4\times 4\times 2$ lattice
with the set of $t$-$J$ model parameters 
that are used in Figs. \ref{spctrm_wc_a}-\ref{spctrm_sc_a}:
$t_1^{\perp} \neq 0$ and $t_2^{\perp} = 0$.
Below, the QCP occurs at $J_{0c} = -1.733 J_1^{\perp}$,
and the integer coordinates $(n_x, n_y)$  specify the 
momentum of the groundstate in units of $\pi/2a$.
And in the case of a doubly degenerate groundstate,
relationships between the
signs of $\langle P_{d\bar d}\rangle$ and of $\langle P_{d\bar d}(\circ)\rangle$ 
are specified by the $\pm$ and $\mp$ symbols.}
\begin{indented}
\item[]\begin{tabular}{@{}lllll}
\br
Observable &
$J_0 = 0$ @ $(0,0)$  & QCP @ $(0,0)$ & QCP @ $(2,0)$ & $J_0 = -23 J_1^{\perp}$ @ $(1,0)$\\
\mr
Fe triplets &
$7.90$ & $11.50$ & $13.48$ & $14.96$\\
$\mu_{\rm hFM}^2 / \mu_{\rm Fe}^2$ &
$0.93$ & $0.60$ & 0.26 & $0.06$\\
$\mu_{\rm cSDW}^2 / \mu_{\rm Fe}^2$ &
$0.08$ & $0.24$ & $0.70$ & $0.94$\\
$P_{d\bar d}$ &
$\pm 1$ & $\pm 1$ & $-1$ & $-1$\\
$P_{d\bar d}(\circ )$ &
$\mp 0.06$ & $\pm 0.02$ & $-0.47$ & $-0.94$\\
\br
\end{tabular}
\end{indented}
\end{table}

\begin{figure}
\includegraphics{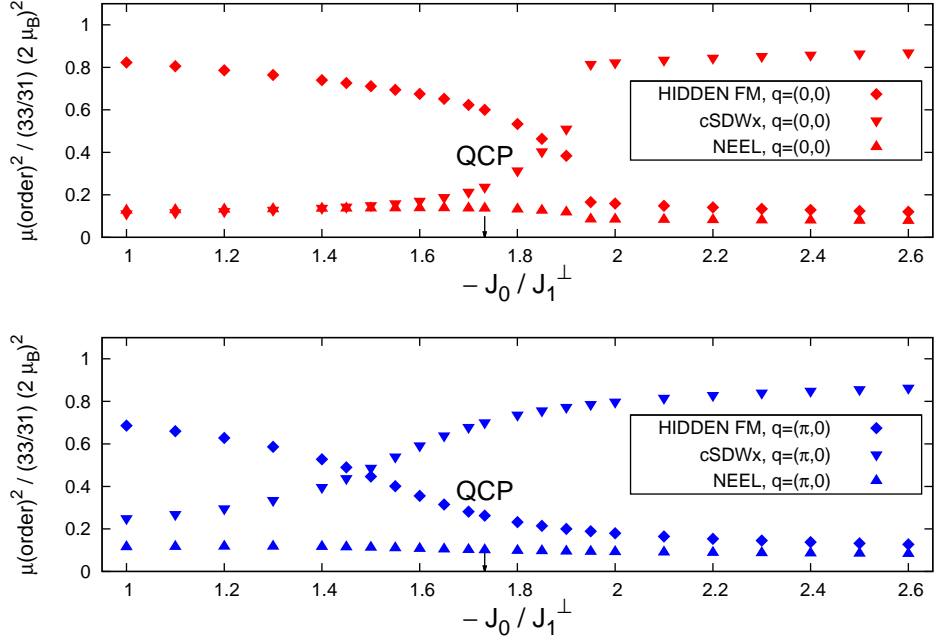}
\caption{Shown is the evolution of the ordered moments about the QCP
for the groundstate at zero (red) and at cSDW (blue) momenta.
Model parameters are listed in the caption to Fig. \ref{spctrm_wc_a}.
The ordered moments for the groundstate at zero 2D momentum (red)
jump at $J_0 = -1.95\, J_1^{\perp}$
because of a level crossing there between the spin-$1/2$ and spin-$3/2$ states. 
(See Fig. \ref{e_vs_j0}, left panel.)}
\label{order_vs_j0}
\end{figure}

\begin{figure}
\includegraphics{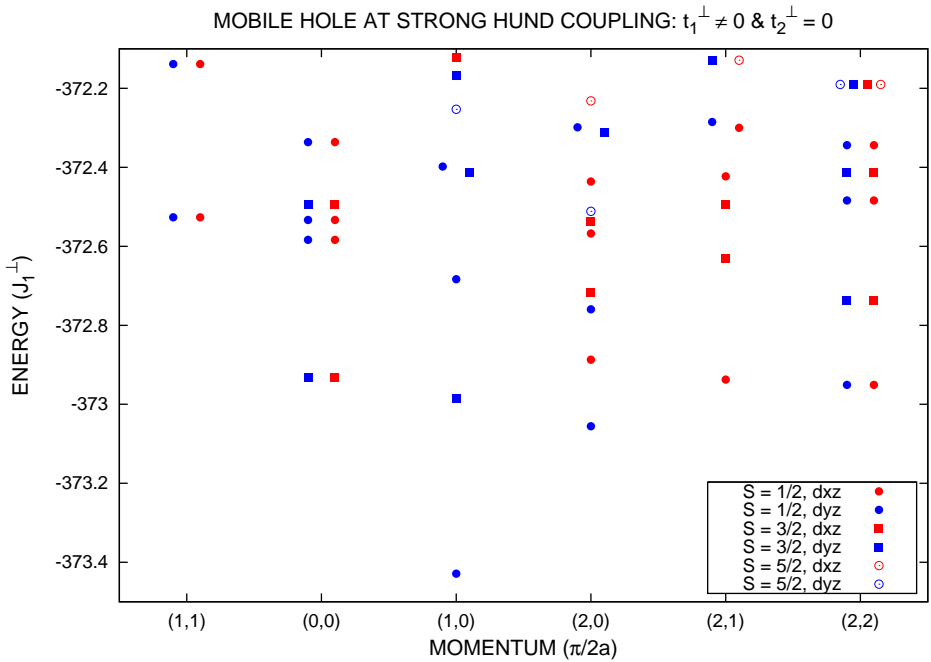}
\caption{The low-energy spectrum of  the two-orbital $t$-$J$ model 
with one mobile hole hopping
over a $4\times 4\times 2$  lattice is displayed at strong Hund's Rule coupling:
$J_0 = -23 J_1^{\perp}$. The remaining parameters are listed in the caption
to Fig. \ref{spctrm_wc_a}.}
\label{spctrm_sc_a}
\end{figure}

Figure  \ref{spctrm_cp_a} shows how the spin-$1/2$ state
at cSDW wavenumber $(\pi / a){\hat{\bi x}}$,
with odd parity under $P_{d\bar d}$,
comes down in energy
with increasing Hund's Rule coupling
to become degenerate with the zero-momentum
doubly-degenerate  groundstate
at a critical coupling
of $J_{0c} = - 1.73 J_1^{\perp}$.
Turning off inter-orbital hopping entirely results in a somewhat higher critical Hund's Rule
coupling of\cite{jpr11} $J_{0c} = - 2.27 J_1^{\perp}$.
We have measured the expectation value for {\it local} orbital swap at the hole iron site,
$P_{d\bar d}(\circ)$.
It has eigenvalues $+1$ and $-1$, 
which correspond to a hole with $3d_{xz}$ and with $3d_{yz}$ orbital character,
respectively.
Table \ref{table1} lists
the corresponding expectation value over the groundstate 
at cSDW wavenumber $(\pi / a){\hat{\bi x}}$:
$\langle P_{d\bar d}(\circ)\rangle = -0.47$.
The emergent hole state therefore  has $74\,\%$ $3d_{yz}$ orbital character. 
Also, the dispersion of the low-energy spin-$1/2$ states shown in Fig. \ref{spctrm_cp_a}
resembles the mean-field prediction for the dispersion of
critical spin-wave excitations shown in Fig. \ref{spin_x}.
The former states are also  well separated from the next excited state at fixed momentum.
This  again is consistent with the  mean-field prediction (\ref{chi_perp}) of well-defined
spin-wave excitations at the QCP.
Finally, the ordered moment
for a hidden ferromagnet (hFM) 
and for a cSDW is defined by
$\mbox{\boldmath$\mu$} (k) = [2\mu_B/(N_{\rm Fe}-{1\over 2})]
 \sum_{\alpha=0}^1 \sum_i e^{i(k_0\alpha + {\bi k}\cdot{\bi r}_i)} {\bi S}_{i,\alpha}$,
with respective 3-momenta
$k = (\pi,0,0)$ and $k = (0,\pi/a,0)$.
Table \ref{table1}
lists the auto-correlation of each
over the groundstate at zero 2D momentum:
$\langle \mbox{\boldmath$\mu$} (k)\cdot \mbox{\boldmath$\mu$} (-k)\rangle_0$.
They are given  in units of the ordered moment of the ferromagnetic state,
$k=(0,0,0)$,
over the $4\times 4\times 2$ lattice:
$\mu_{\rm Fe}^2 = (33/31)(2\mu_B)^2$.
Notice
that $\mu_{cSDW}^2$ remains small at the QCP in the case of the zero-momentum groundstate,
which is consistent with neutron diffraction studies
in iron-pnictide systems\cite{delacruz}.
Notice also
that $\mu_{hFM}^2$ remains sizable at the quantum critical point
for the zero-momentum groundstate,
which indicates that hidden half-metal character persists there.
Figure \ref{order_vs_j0} displays the evolution of these ordered moments near the QCP.
The QCP bisects the points at
which $\mu_{hFM}^2$ and $\mu_{cSDW}^2$ dovetail for
groundstates with 2D momenta at zero and at $(\pi/a)\hat{\bi x}$.
Also displayed is the square of the moment $\mu(k)$ for N\' eel order, $k = (0,\pi/a,\pi/a)$,
which always remains small.
The large values of $\mu_{hFM}^2$ displayed by Fig. \ref{order_vs_j0} at Hund coupling below critical
in conjunction with the mean-field prediction, Eqs. (\ref{chi_perp}) and (\ref{spctrl_wght}),
argues strongly for the persistence of hidden ferromagnetic order in the thermodynamic limit.
With the exception of the QCP,
theoretical predictions for long-range cSDW order in the two-orbital
$t$-$J$ model (\ref{tJ}) are absent, however.  (Cf. section 5.)
It is  therefore unclear whether the cSDW order displayed by Fig. \ref{order_vs_j0} 
at Hund coupling above critical survives the thermodynamic limit.


Last, 
Fig. \ref{spctrm_sc_a} displays the exact low-energy spectrum for a $4\times 4$ lattice of
true spin-$1$ iron atoms with one mobile hole. In particular, 
Hund's Rule is enforced by setting $J_0 = -23 J_1^{\perp}$.
As before, red points and blue points are even and odd under global orbital swap, $P_{d\bar d}$.
We measured the expectation values for local orbital swap at the hole iron site, $P_{d\bar d}(\circ)$,
and we found that the hole in even and odd parity states has orbital character that is
over $95\,\%$ $3d_{xz}$ and $3d_{yz}$, respectively .
(See Table \ref{table1}.)
We have therefore replaced the former labels with the latter ones in the legend to Fig.  \ref{spctrm_sc_a}.
The groundstate notably has spin-$1/2$,
it carries momentum $(\pi / 2a){\hat{\bi x}} ({\hat{\bi y}})$,
and it has orbital $3d_{yz} (3d_{xz})$ character. 
This is analogous to the groundstate momentum of
$(\pi / 2a)({\hat{\bi x}} \pm {\hat{\bi y}})$
that is predicted for one mobile hole in a 2D N\' eel state
by Kane, Lee and Read\cite{kane89}.
Table \ref{table1} lists ordered moments computed in the groundstate
at momentum $(\pi / 2a){\hat{\bi x}} ({\hat{\bi y}})$.
These moments combined with the low-energy spectrum suggest
that the groundstate
at thermodynamic hole densities $x$ is  a robust cSDW metal
with Fermi surfaces that are  centered at wavenumbers
$(\pi / 2a){\hat{\bi x}}$ and  $(\pi / 2a){\hat{\bi y}}$.
This state is therefore unable to account for
the Fermi surfaces that are  observed experimentally in iron-pnictide systems.

\begin{table}
\caption{\label{table2} Listed are the characteristic properties of observable versus
hidden quantum-critical spin-wave excitations in the two-orbital Heisenberg model 
that corresponds to the two-orbital $t$-$J$ model (\ref{tJ}) 
at half filling.
(See ref. \cite{jpr10}, Fig. 5.)}
\begin{indented}
\item[]\begin{tabular}{@{}lllll}
\br
spin wave & order parameter & 2D momentum & Hund's Rule? & $P_{d\bar d}$ \& $P_{d\bar d}^{\prime}$ \\
\mr
observable & ${\bi S}_{i,d-} + {\bi S}_{i,d+}$ 
& $(\pi/a){\hat{\bi x}} ({\hat{\bi y}})$ & obeyed & even\\
hidden & ${\bi S}_{i,d-} - {\bi S}_{i,d+}$ & zero & violated & odd\\
\br
\end{tabular}
\end{indented}
\end{table}
\begin{figure}
\includegraphics{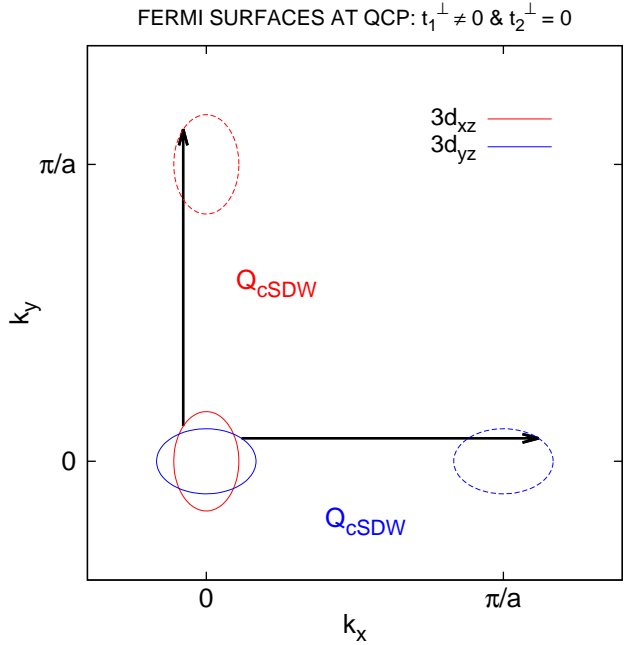}
\caption{Emergent nesting of Fermi surface pockets at QCP 
with nearest-neighbor inter-orbital
hopping: negative $t_1^{\perp}(\hat{\bi x}) = - t_1^{\perp}(\hat{\bi y})$ and $t_2^{\perp} = 0$.  
(Cf. figs. \ref{FS} and \ref{spctrm_cp_a}.)}
\label{conjecture_a}
\end{figure}

Figure \ref{order_vs_j0} provides evidence for a QCP
that separates a cSDW metal from a hidden half metal
as Hund coupling gets weaker.
The exact quantum-critical spectrum
displayed by Fig. \ref{spctrm_cp_a}
reveals 
groundstates at zero 2D momentum
that are degenerate with groundstates
at cSDW momenta $(\pi/a)\hat {\bi x}$ and $(\pi/a)\hat {\bi y}$.
The  expectation values for  $\langle P_{d\bar d}(\circ)\rangle$
that are listed in Table \ref{table1} 
imply that the latter are emergent mobile hole states with
$74\,\%$ $3d_{yz}$ and $74\,\%$ $3d_{xz}$ orbital character, respectively. 
Coherent inter-orbital hole propagation at cSDW wavelengths
therefore exists at the QCP
in the case where inter-orbital hopping is purely across nearest neighbors.
This contrasts with the prediction of dynamical suppression of inter-orbital hopping
by the previous Schwinger-boson-slave-fermion mean-field theory at large $s_0$.
General agreement between the two calculations nevertheless exists.
In particular,
both the exact results shown by Fig. \ref{spctrm_cp_a}
and the previous Schwinger-boson-slave-fermion
mean-field theory analysis find a QCP at moderate Hund's Rule coupling, $-J_0$,
where one-hole groundstates at zero 2D momentum and at cSDW momenta become degenerate.
The mean-field theory predicts  that 
the momentum  of the degenerate groundstate in Fig. \ref{spctrm_cp_a}
is carried entirely
by a cSDW spinwave that softens to zero excitation energy at the QCP.
(See Fig. \ref{spin_x}.)
Table \ref{table2} summarizes the nature of quantum-critical spinwave excitations
for the corresponding Heisenberg model in the absence of mobile holes\cite{jpr10}.
The key point to notice is that quantum-critical spinwaves
at cSDW momenta are {\it observable},
with even parity under $P_{d\bar d}$,
while those at zero 2D momentum are {\it hidden},
with odd parity under $P_{d\bar d}$. 
[See Eqs. (\ref{chi_perp}) and (\ref{spctrl_wght}), and Fig. \ref{spin_x}.]
We have confirmed this  by exact diagonalization of the corresponding Heisenberg model.
Combining the mean-field-theory picture of the QCP with the present exact results then
leads to the following conjecture:
the critical cSDW spinwave at
wavenumber $(\pi / a){\hat{\bi x}}$ relates
the odd parity $3d_{yz}$ portion of the hole pockets centered at zero 2D momentum (Fig. \ref{FS}) 
with an emergent Fermi-surface pocket centered at $(\pi / a){\hat{\bi x}}$
that has the same orbital character,
while the critical cSDW spinwave at
wavenumber $(\pi / a){\hat{\bi y}}$ 
relates the even parity $3d_{xz}$ portion of the hole pockets centered at zero 2D momentum 
with an emergent Fermi-surface pocket centered at $(\pi / a){\hat{\bi y}}$
that has the same orbital character.
Figure \ref{conjecture_a} summarizes this emergent nesting mechanism.
Notice that Fermi surface pockets centered at cSDW momenta 
have orbital character that is purely $3d_{xz}$ and $3d_{yz}$,
with no admixture of other $3d$ orbitals.
This is consistent with a recent determination of the electronic structure in
optimally doped Ba(Fe$_{1-x}$Co$_x$)$_2$As$_2$ by ARPES\cite{zhang11}. 

\begin{figure} 
\includegraphics{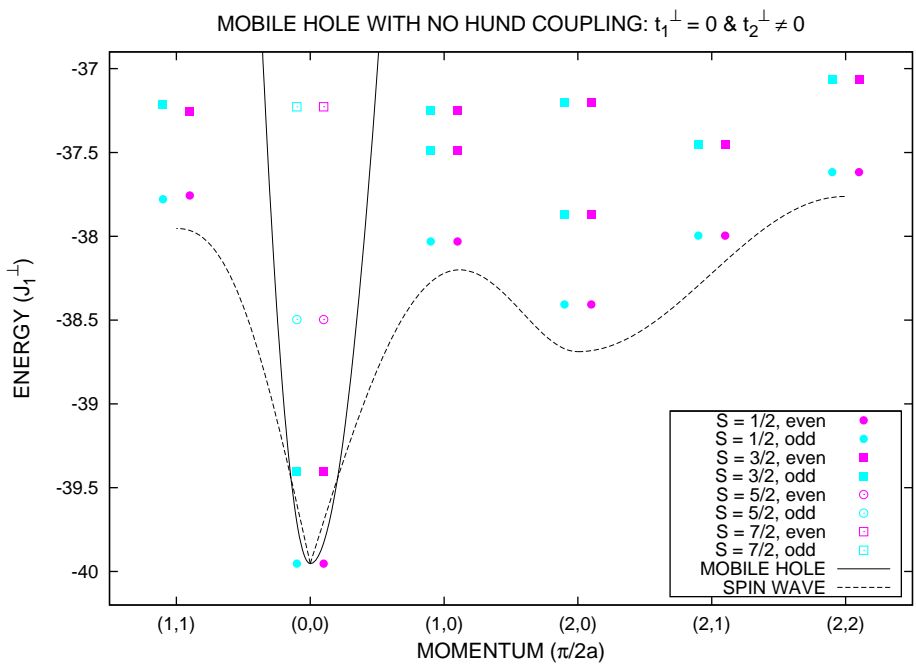}
\caption{Shown is the low-energy spectrum of the two-orbital $t$-$J$ model,
Eq. (\ref{tJ}),  over a
$4\times 4\times 2$  lattice with one hole,
in the absence of Hund's Rule: $J_0 = 0$.
The remaining parameters are
$J_1^{\parallel} = 0$,
$J_1^{\perp} > 0$,
$J_2^{\parallel} = 0.3 J_1^{\perp} = J_2^{\perp}$,
$t_1^{\parallel} = - 5 J_1^{\perp}$,
$t_1^{\perp} = 0$,
$t_2^{\parallel} = 0$,
$t_2^{\perp} ({\hat{\bi x}}+{\hat{\bi y}}) = - i J_1^{\perp}$, 
and $t_2^{\perp} (-{\hat{\bi x}}+{\hat{\bi y}}) = + i J_1^{\perp}$.
Pink and light blue states are respectively even and odd under $P_{d\bar d}^{\prime}$.
A comparison with the hole spectrum, 
$\varepsilon_f(k) = \varepsilon_{\parallel}({\bi k})$, 
and with the spin-wave spectrum,
$\omega_b(k) = \omega_b^{(0)}({\bi k})$,
at large $s_0$ and $x=0$ is also shown.}
\label{spctrm_wc_b}
\end{figure}
\begin{figure}
\includegraphics{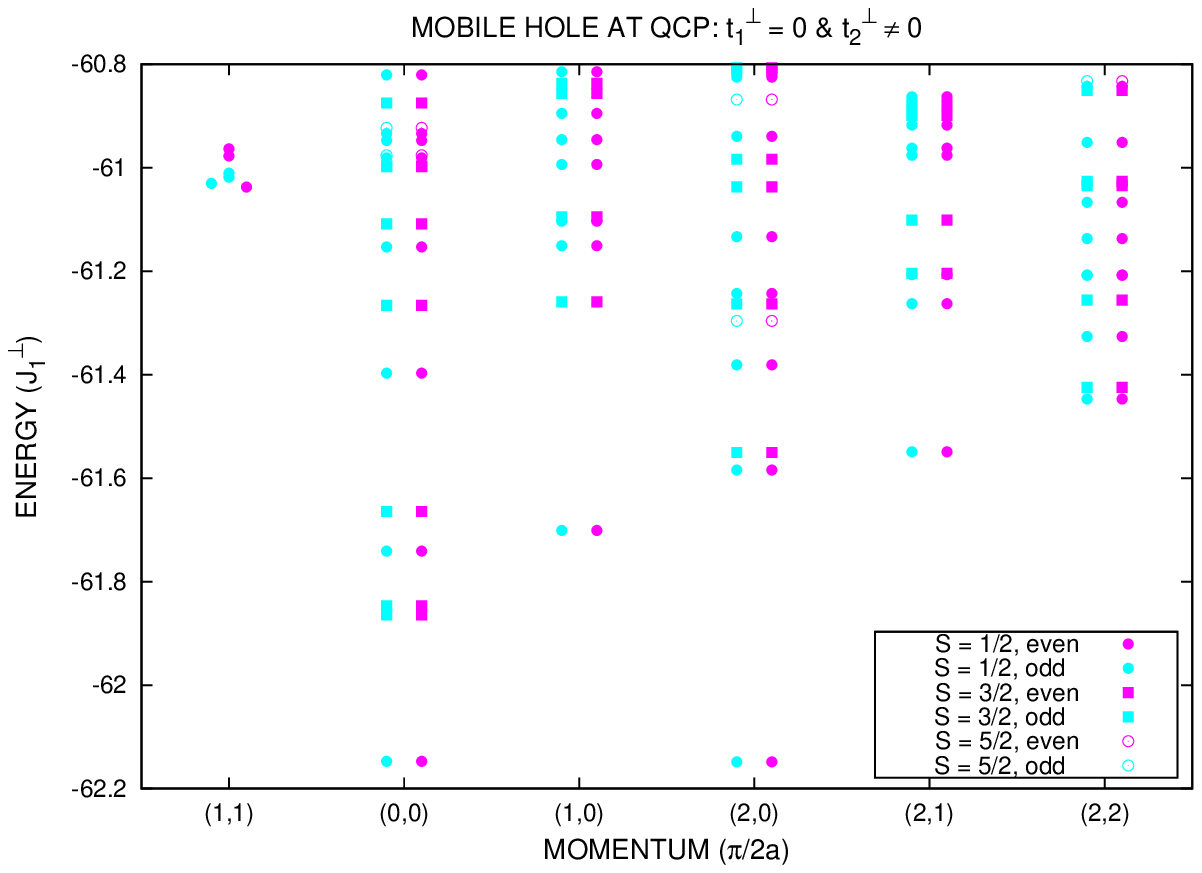}
\caption{The low-energy spectrum of the two-orbital $t$-$J$ model with one hole roaming
over a $4\times 4\times 2$  lattice is displayed at the QCP that separates the cSDW from the
hidden half-metal state: $J_{0c} = -2.26 J_1^{\perp}$. The remaining set of parameters are listed
in the caption to the Fig. \ref{spctrm_wc_b}.
Pink and light blue states are respectively even and odd under $P_{d\bar d}^{\prime}$.}
\label{spctrm_cp_b}
\end{figure}

Figures \ref{spctrm_wc_b} and \ref{spctrm_cp_b} show the exact spectrum of the 
two-orbital $t$-$J$ model (\ref{tJ}),
respectively, in the absence of Hund's Rule coupling and at the QCP,
but at the other extreme where inter-orbital hopping
is purely across next-nearest neighbors.
The model parameters, in particular, are
$J_1^{\parallel} = 0$,
$J_1^{\perp} > 0$,
$J_2^{\parallel} = 0.3 J_1^{\perp} = J_2^{\perp}$,
$t_1^{\parallel} = - 5 J_1^{\perp}$,
$t_1^{\perp} = 0$,
$t_2^{\parallel} = 0$,
$t_2^{\perp} ({\hat{\bi x}}+{\hat{\bi y}}) = - i J_1^{\perp}$,
and $t_2^{\perp} (-{\hat{\bi x}}+{\hat{\bi y}}) = + i J_1^{\perp}$.
Pink and light blue states are respectively 
even and odd under the symmetry operation $P_{d\bar d}^{\prime}$
that  denotes swap of the $d\pm$ orbitals after the gauge transformation $e^{\pm i \pi/4}$.
The Hund's Rule coupling at the QCP is $-J_{0c} = 2.26 J_1^{\perp}$,
which is very close to its value in the absence of inter-orbital hopping\cite{jpr11},
$-J_{0c} = 2.27 J_1^{\perp}$.  
Notice that both values are almost three times
larger than the mean-field prediction (\ref{J_0c}) at $x=0$,  $-J_{0c} = 0.8 J_1^{\perp}$.
The quantum-critical spectrum shown by
Fig. \ref{spctrm_cp_b} is also very close to the corresponding one in the absence of inter-orbital
hopping up to a rigid energy shift that is relatively small\cite{jpr11}.
Further, the moments for hidden ferromagnetic order and for cSDW order at the QCP
that are listed in Table \ref{table3} match those obtained previously
in the absence of inter-orbital hopping\cite{jpr11} to within $1 \%$.
We also computed the groundstate expectation values of modified orbital swap at the iron hole site, 
$P_{d\bar d}^{\prime}(\circ)$, and these are listed in Table \ref{table3}.
A hole in a $3d_{x^{\prime} z}$
 orbital has even parity ($+1$) under it,
while a hole in a $3d_{y^{\prime} z}$
 orbital has odd parity ($-1$) under it.
Here, $x^{\prime} = (x+y)/2^{1/2}$ and $y^{\prime} = (y-x)/2^{1/2}$
are the 2D coordinates along the next-nearest-neighbor links.
Notice that $\langle P_{d\bar d}^{\prime}(\circ) \rangle$ is generally small compared to unity, 
which means that the hole does {\it not} 
possess well-defined $3d_{x^{\prime} z}$ or $3d_{y^{\prime} z}$
 orbital character at the QCP.

To conclude, good agreement exists between exact results for 
the spectrum of the hidden half metal 
in the presence of purely next-nearest-neighbor inter-orbital hopping 
and dynamical suppression of the latter,
which is predicted by the previous Schwinger-boson-slave-fermion mean-field theory
at large electron spin $s_0$.
The two-fold degeneracy of the groundstate at cSDW momenta indicates that
it is a special case, however.
On the contrary,
although the former case of purely nearest-neighbor inter-orbital hopping is ideal,
we believe that it is representative of the general case 
where both types of inter-orbital hopping are present. (See Fig. \ref{FS}.)


%
\begin{table}
\caption{\label{table3} 
Listed are groundstate expectation values of physical observables
for one hole hopping over a $4\times 4\times 2$ lattice
with the set of $t$-$J$ model parameters
that are used in Figs. \ref{spctrm_wc_b} and \ref{spctrm_cp_b}:
$t_1^{\perp} = 0$ and $t_2^{\perp} \neq 0$.
Below, the QCP occurs at $J_{0c} = -2.26 J_1^{\perp}$,
and the integer coordinates $(n_x, n_y)$  specify 
the momentum of the groundstate in units of $\pi/2a$.
Relationships between the
signs of $\langle P_{d\bar d}\rangle$ and of $\langle P_{d\bar d}(\circ)\rangle$ 
are specified by the $\pm$ and $\mp$ symbols
for these doubly degenerate groundstates.}
\begin{indented}
\item[]\begin{tabular}{@{}llll}
\br
Observable &
$J_0 = 0$ @ $(0,0)$  & QCP @ $(0,0)$ & QCP @ $(2,0)$\\
\mr
Fe triplets &
$7.84$ & $12.09$ & $13.20$\\
$\mu_{\rm hFM}^2 / \mu_{\rm Fe}^2$ &
$0.94$ & $0.53$ & 0.32\\
$\mu_{\rm cSDW}^2 / \mu_{\rm Fe}^2$ &
$0.08$ & $0.20$ & $0.62$\\
$P_{d\bar d}^{\prime}$ &
$\pm 1$ & $\pm 1$ & $\pm 1$\\
$P_{d\bar d}^{\prime}(\circ )$ &
$\mp 0.07$ & $\mp 0.06$ & $\pm 0.13$\\
\br
\end{tabular}
\end{indented}
\end{table}

\section{Normal State and Spin Fluctuations}
We now propose a normal metallic groundstate for the 
two-orbital $t$-$J$ model (\ref{tJ}) 
that exhibits a low density of charge carriers and cSDW nesting.
Ultimately, we will argue for electron-type dispersion of 
the nested bands that emerge near cSDW momenta 
in mean-field theory and in exact results for one hole.
(See section 3.3 and Fig. \ref{spctrm_cp_a}).
The cSDW-metal groundstate shall be constructed in two stages.

\begin{figure}
\includegraphics{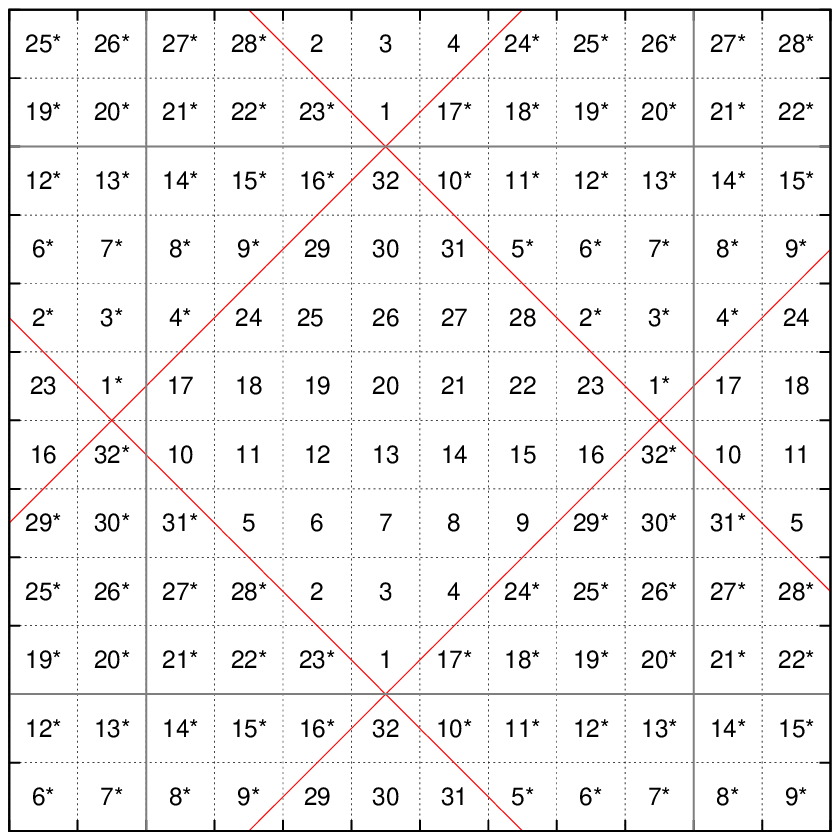}
\caption{Shown is an $8\times 8$ square lattice of iron atoms under periodic boundary conditions
that is split into two tilted squares. (See ref. \cite{oitmaa_betts_78}.)
The unique site that is farthest away from site $i$
is labeled by $i^*$.}
\label{split_square}
\end{figure}

The first piece of the new cSDW metal state exhibits low-energy spin fluctuations in the hidden channel
${\bi S}_{i,d-} - {\bi S}_{i,d+}$ at 2D momentum ${\bi q} = 0$.
It is obtained after a Gutzwiller projection\cite{Gutz} of the corresponding Fermi gas,
\begin{equation}
|\Phi_0(x_+,0)\rangle = \prod_{k_0,{\bi k},s}^{\qquad\prime}
c_s(k_0,{\bi k})^{\dagger}|0\rangle \quad{\rm with}\quad
 \varepsilon_e(k_0,{\bi k}) < \varepsilon_+,
\label{Phi_00}
\end{equation}
in which case all interaction terms are suppressed,
but where one-electron states are restricted to ``black squares'' of the ``checkerboard''
in momentum space:
${\bi k} = (2\pi n_x / N a, 2\pi n_y / N a)$, 
with integers $n_x$ and $n_y$ either both even or both odd,
and with even integer $N$.
The electron annihilation operator above has the form
\begin{equation}
c_s (k_0,{\bi k}) =
{e^{+i\delta_e({\bi k})}\over{2^{1/2}}}
c_{d-,s}({\bi k}) +
e^{ik_0} {e^{-i\delta_e({\bi k})}\over{2^{1/2}}}
c_{d+,s}({\bi k}) ,
\label{c_s}
\end{equation}
where $c_{\alpha,s}({\bi k}) = N_{\rm Fe}^{-1/2}\sum_i e^{-i{\bi k}\cdot{\bi r_i}} c_{i,\alpha,s}$.
Also, the phase shift and energy eigenvalues of one-electron states are given by
$\delta_e({\bi k}) = - \delta_f ({\bi k})$
and
$\varepsilon_e(k_0,{\bi k}) = -\varepsilon_f(k_0,{\bi k})$.  
Here, the mean-field parameters
$Q_n^{\parallel}$ and $R_n^{\perp}$ that appear in the expressions for the matrix elements
$\varepsilon_{\parallel}$ (\ref{epsilon_para}) and $\varepsilon_{\perp}$ (\ref{epsilon_perp})
are instead all set to $s_0 = 1/2$.
Periodic boundary conditions are assumed over a square lattice of iron atoms.
It has dimensions $N a\times N a$, with $N_{\rm Fe} = N^2$ iron atoms.
The one-electron states occupied by the electron gas (\ref{Phi_00}) are then also periodic 
when translated  to the farthest point away, in which case the translation
vector is ${\bi L}_{*} = (N/2) a (\hat{\bi x}\pm\hat{\bi y})$.
It splits the square region into two tilted subsquares,
as shown by Fig. \ref{split_square}.
In particular,
if iron sites $j$ are restricted to one of the subsquares,
and if $j^*$ is the site that is farthest away from it, 
${\bi r}_{j^*} = {\bi r}_j + {\bi L}_*$,
then the one-electron states with momenta on ``black squares'' of the ``checkerboard''
have the form
$\langle \alpha,i|c_s(k_0,{\bi k})^{\dagger}|0\rangle 
= \phi_{k_0,{\bi k}}(\alpha,i) \chi_s$,
where
\begin{equation}
\phi_{k_0,{\bi k}}(\alpha,i) = \Biggl[
{e^{-i\delta_e({\bi k})}\over{2^{1/2}}} \delta_{\alpha,d-}
+ e^{ik_0} {e^{+i\delta_e({\bi k})}\over{2^{1/2}}} \delta_{\alpha,d+}\Biggr]
\sum_{j}{^\prime} {e^{i{\bi k}\cdot{\bi r_j}}\over{N_{\rm Fe}^{1/2}}} (\delta_{i,j}+\delta_{i,j^*}) ,
\label{1e_stts}
\end{equation}
and where $\chi_s$ represents spin $s=\uparrow , \downarrow$.
Notice that filling both the bonding and the anti-bonding bands,
$k_0 = 0$ and $\pi$, 
within the restricted momentum space
corresponds to a band insulator with ${\cal N} = 2 N_{\rm Fe}$ electrons
that are restricted to one of the tilted subsquares in Fig. \ref{split_square}.
Setting the Fermi level $\varepsilon_+$ just below the top of the bands 
yields two hole pockets centered at zero 2D momentum, on the other hand,
which are depicted by Fig. \ref{FS}.
We now introduce
the following Gutzwiller projection\cite{Gutz}
of the Fermi gas (\ref{Phi_00})
as a candidate metallic state of the two-orbital $t$-$J$ model:
\begin{equation}
|\Psi_0\rangle = \prod_{i,\alpha} {\rm exp}(-g_0 n_{i,\alpha,\uparrow} n_{i,\alpha,\downarrow})
                 \prod_j {\rm exp}(-g_0^{\prime} n_{j,d+} n_{j,d-}) |\Phi_0\rangle ,
\label{Psi_0}
\end{equation}
where  $g_0 \rightarrow \infty$.  This choice for one of the variational parameters
strictly prohibits double occupancy per site-orbital. 
Last, notice that the arguments of the exponentials above are invariant under spin rotations.
The proposed Gutzwiller-projected state (\ref{Psi_0}) therefore inherits the spin-singlet nature of the
Fermi gas (\ref{Phi_00}).

It is instructive to compute the above Gutzwiller wavefunction
in the case where the Fermi level $\varepsilon_+$ lies above the maximum of both bands
at ${\bi k} = 0$.
The Fermi gas (\ref{Phi_00}) is then a Slater determinant over all of the allowed
one-electron states restricted to ``black squares'' of the ``checkerboard'' in 2D momentum.  
It is evaluated directly in Appendix B, Eq. (\ref{slater_deter_intrmdt}).
The Gutzwiller projection of it (\ref{Psi_0}) is a product of spin-singlet pairs within
the $d+$ and $d-$ orbitals that entangle opposite spins separated by the maximum displacement ${\bi L}_*$:
\begin{eqnarray}
\langle 1,\ldots,{\cal N}|\Psi_0\rangle = 
{1\over{\sqrt{2^{N_{\rm Fe}}{\cal N}!}}} \sum_p ({\rm sgn}\, p) \prod_{m=1}^{N_{\rm Fe}}
& {1\over \sqrt 2}
(\uparrow_{p(m)}\downarrow_{p(m^{\prime})}-\downarrow_{p(m)}\uparrow_{p(m^{\prime})})\cdot \nonumber \\
& \cdot \langle p(m), p(m^{\prime})|\alpha_m,i_m; \alpha_m,i_m^*\rangle_+  , \nonumber \\
\label{slater_deter_half_fill}
\end{eqnarray}
where $m^{\prime} = m + N_{\rm Fe}$, and where
$|\alpha,i; \beta,j\rangle_+ = 
(|\alpha,i\rangle |\beta,j\rangle + 
|\beta,j\rangle |\alpha,i\rangle)/\sqrt 2$.
Here $(\alpha_1,i_1), (\alpha_2, i_2), \ldots , (\alpha_{N_{\rm Fe}}, i_{N_{\rm Fe}})$
is a list of all orbitals and of 
all sites within a single subsquare in Fig. \ref{split_square}.
Above, $|\Psi_0\rangle$ describes a featureless paramagnetic insulator in a spin-singlet state:
$\langle {\bi S}_{i,\alpha} \cdot {\bi S}_{j,\beta}\rangle = 0$ for $j\neq i,i^*$
or for $\beta\neq \alpha$,
and  $({\bi S}_{i,\alpha} + {\bi S}_{i^*,\alpha}) |\Psi_0\rangle = 0$.
The former yields an average number of triplets per iron site of
$\langle ({\bi S}_{i,d+} +  {\bi S}_{i,d-})^2 \rangle / 2\hbar^2 = 3/4$.
It lies halfway in between the value of $1/2$ for the naive hidden ferromagnetic state
with perfect $\nwarrow_{d+}\searrow_{d-}$ spin order
and the maximum value of $1$ when Hund's Rule is obeyed.
The value $3/4$ also coincides with the probability that a given iron atom be found
in a spin-$1$ state.
It must also be mentioned that the featureless paramagnetic insulator
(\ref{slater_deter_half_fill}) is translationally invariant
and $4$-fold rotationally invariant.

The paramagnetic nature of the Gutzwiller wave function (\ref{Psi_0}) 
at half filling (\ref{slater_deter_half_fill})
suggests that it describes a normal metallic phase
when the Fermi level $\varepsilon_+$ falls below the band maxima at zero 2D momentum
in the Fermi gas state (\ref{Phi_00}).
Evaluating it becomes non-trivial then, however.
We can get some idea of the nature of 
long-wavelength spin correlations
in the Gutzwiller wave function (\ref{Psi_0}) away from half filling
by artificially turning off all of the interactions
in the two-orbital $t$-$J$ model (\ref{tJ}).  
This leaves us with the Fermi gas state (\ref{Phi_00})
and with the hopping terms in the Hamiltonian.
It results in
two hole Fermi surface pockets
centered at zero 2D momentum,
which are depicted by Fig. \ref{FS}.
We can then compute the dynamical auto-correlation
of the true spin ($+$) and of the hidden spin ($-$) at the longwavelength limit,
$N_{\rm Fe}^{-1/2} \sum_i ({\bi S}_{i,d-} \pm {\bi S}_{i,d+})$,
within this approximation.  
Explicit Lindhard functions
for the dynamical spin-spin auto-correlation function 
in multi-orbital Fermi gases are given in ref. \cite{graser_09}.
Substituting the matrix elements for the one-electron states (\ref{1e_stts})
into those expressions yields the following results
at the long wavelength limit:
$\langle {\bi S}(+)\cdot {\bi S}^{\prime}(+)\rangle |_{0,\omega}  = 0$ and
%
\begin{equation}
i\langle {\bi S}(-)\cdot {\bi S}^{\prime}(-)\rangle |_{0,\omega}^{\perp}  = 
{\hbar^2\over{N_{\rm Fe}}} \sum_{k_0}\sum_{\bi k}^{\qquad\prime}{n_F(k_0,{\bi k}) - n_F(k_0+\pi,{\bi k})\over
{\omega+\varepsilon_e(k_0+\pi,{\bi k})-\varepsilon_e(k_0,{\bi k})+i0^+}} .
\label{Lindhard}
\end{equation}
Above, the symbol $\perp$ indicates that the dot product is restricted to the $x$ and $y$
components of the spin,
the prime notation 
indicates that the summation over 2D momentum  is restricted to
the ``black squares'' of the ``checkerboard'', and
$n_F(k) = \theta[\varepsilon_+-\varepsilon_e(k)]$ at zero temperature.
True spin fluctuations at ${\bi q} = 0$ are therefore absent. 
On the other hand,
at zero temperature and $\omega > 0$,
the imaginary part of the above expression
for hidden spin fluctuations at ${\bi q} = 0$
reduces to
\begin{equation}
{\rm Re}\, \langle {\bi S}(-)\cdot {\bi S}^{\prime}(-)\rangle |_{0,\omega}^{\perp} =
  {\pi\hbar^2\over{N_{\rm Fe}}} \sum_{\bi k}^{\qquad\prime}
n_F(0,{\bi k})[1-n_F(\pi,{\bi k})]
\delta[\omega - 2|\varepsilon_{\perp}({\bi k})|] .
\label{hidden_spin_fluct}
\end{equation}
The particle-hole occupation factors in the sum above
restrict momenta to lie in between the two hole-pocket Fermi surfaces 
that are shown in Fig. \ref{FS}.
The present approximation therefore predicts hidden spin fluctuations at long wavelength,
at frequencies inside of the range $[\omega_1,\omega_2]$, with
$\omega_1 = 2\, {\rm min}\, |\varepsilon_{\perp}|$ 
and
$\omega_2 = 2\, {\rm max}\, |\varepsilon_{\perp}|$
for momenta restricted to lie in between the two hole pockets.
At low hole concentration, $x_+ \ll 1$,
and at low inter-orbital hopping, $|t_{\perp}|,|t_{\perp}^{\prime}|\ll -t_{\parallel}$,
Eqs. (\ref{FSS}) and (\ref{e_f_vs_x}) for the hole-pocket Fermi surfaces yield that
the limits in frequency are approximately
$\omega_{1} \cong 8\pi x_+ \, {\rm min}\, ([1 - |t_{\perp}|/(-t_{\parallel})]|t_{\perp}|,
 [1 - |t_{\perp}^{\prime}|/(-t_{\parallel})]|t_{\perp}^{\prime}|)$
and
$\omega_2 \cong 8\pi x_+ \, {\rm max}\, ([1 + |t_{\perp}|/(-t_{\parallel})]|t_{\perp}|,
 [1 + |t_{\perp}^{\prime}|/(-t_{\parallel})] |t_{\perp}^{\prime}|)$.
Expression (\ref{hidden_spin_fluct}) also  yields that 
the integrated spectral weight of the hidden spin fluctuations,
$A(-,0) = \int_0^{\infty} d\omega\, 
{\rm Re}\, \langle {\bi S}(-)\cdot {\bi S}^{\prime}(-)\rangle |_{0,\omega}^{\perp}$,
is proportional to the fraction of momentum space that lies in between the two hole pockets
shown in Fig. \ref{FS}:
$A(-,0) =  (\pi\hbar^2/N_{\rm Fe}) \sum_{\bi k}^{\prime} n_F(0,{\bi k})[1-n_F(\pi,{\bi k})]$.
It is of order 
$A(-,0)/\hbar^2\sim  x_+ \, {\rm max}(|t_{\perp}|,|t_{\perp}^{\prime}|) / (-t_{\parallel})$
within the present limits.
This estimate for the spectral weight of hidden spin fluctuations at the long wavelength limit
must be compared to the divergent spectral weight (\ref{spctrl_wght}) predicted by
mean-field theory for the hidden half-metal state.
The range of hidden magnetic correlations is finite in the former case,
whereas it is infinite in the latter case.

Recall the exact energy spectrum
of one hole in the $t$-$J$ model (\ref{tJ})
displayed by Fig.  \ref{spctrm_cp_a},
at Hund's Rule coupling of moderate strength,
and in the absence of inter-orbital next-nearest-neighbor hopping,
$t_2^{\perp} = 0$.
Low-energy electronic states emerge at cSDW momenta with specific orbital character.
It suggests the following modification to the Fermi-gas part (\ref{Phi_00})
of the variational state (\ref{Psi_0})
that adds emergent nesting of Fermi surfaces at cSDW momenta:
\begin{equation}
|\Phi_0(x_+,x_-)\rangle = \prod_{k_0,{\bi k},s}^{\qquad\prime}
c_s(\pi,{\bi k+{\bi Q_{k}}})^{\dagger}|\Phi_0(x_+, 0)\rangle \quad{\rm with}\quad
 \varepsilon_e(k_0,{\bi k}) > \varepsilon_-,
\label{Phi_0}
\end{equation}
where ${\bi Q}_{k_0,{\bi k}} = (\pi/a){\hat{\bi x}}$
if $P_{\bi Q}(k_0,{\bi k};\pi,{\bi k}+{\bi Q}) > 1/2$,
and where ${\bi Q}_{k_0,{\bi k}} = (\pi/a){\hat{\bi y}}$ 
if $P_{\bi Q}(k_0,{\bi k};\pi,{\bi k}+{\bi Q}) > 1/2$.
Here, 
$P_{\bi Q}(k;k^{\prime}) = 
|\sum_{\alpha}\sum_{i} \phi_{k}^*(\alpha,i) e^{-i{\bi Q}\cdot{\bi r}_i} \phi_{k^{\prime}}(\alpha,i)|^2$
is the nesting probability between two eigenstates of the hopping Hamiltonian for free electrons.
Substituting in the form (\ref{1e_stts}) for the eigenstates yields
$P_{\bi Q}(\pi,{\bi k};\pi,{\bi k}+{\bi Q}) = {\rm cos}^2[\delta_e({\bi k})-\delta_e({\bi k}+{\bi Q})]$
and $P_{\bi Q}(0,{\bi k};\pi,{\bi k}+{\bi Q}) = {\rm sin}^2[\delta_e({\bi k})-\delta_e({\bi k}+{\bi Q})]$.
The above variational state (\ref{Phi_0}) requires
unoccupied one-electron states
that carry nested $3$-momenta $(\pi,{\bi k}+{\bi Q}_{k_0,{\bi k}})$
in the Fermi gas $\Phi_0(x_+,0)$ (\ref{Phi_00}).
Hence, $N/2$ must be odd.
The nested Fermi surfaces of the Fermi gas (\ref{Phi_0})
are displayed by Fig. \ref{conjecture_a}
for negative $t_1^{\perp}(\hat{\bi x}) = -t_1^{\perp}(\hat{\bi y})$ 
and $t_2^{\perp} = 0$.
In the general case with hybridization,
where
$t_2^{\perp}(\hat{\bi x}+\hat{\bi y}) = - t_2^{\perp}(-\hat{\bi x}+\hat{\bi y})$
is pure imaginary,
Eq. (\ref{epsilon_perp_bis}) for the inter-orbital hopping matrix element
yields that $\delta_e({\bi k}) \cong 0$ for ${\bi k}$ near $(\pi/a)\hat{\bi x}$
and that $\delta_e({\bi k}) \cong \pi/2$ for ${\bi k}$ near $(\pi/a)\hat{\bi y}$.
Also, $\delta_e({\bi k})$ winds around the hole pockets following (\ref{phase_factor_at_0})
for ${\bi k}$ near zero 2D momentum. 
In the general case,
the pattern of nesting defined above is then simply that described by Fig. \ref{conjecture_a},
but with hybridization between the $3d_{xz}$ and $3d_{yz}$ orbitals included.

At perfect nesting,  $\varepsilon_+ = \varepsilon_-$,
the Gutzwiller projection (\ref{Psi_0}) of the Slater determinant (\ref{Phi_0})
is necessarily a Mott insulator at half filling: $x_+ = x_-$.  
It is therefore a variational groundstate for the corresponding 
two-orbital Heisenberg model over the square lattice.
Both exact and spin-wave analyses of that model
finds evidence for a quantum-critical transition between
a hidden ferromagnetic state at weak Hund coupling and a cSDW at strong Hund coupling\cite{jpr10}.
(See Fig. \ref{phase_diagram}.)
The perfect nesting of the Fermi gas (\ref{Phi_0}) at half filling suggests, however, 
that its Gutzwiller projection $\Psi_0$
shows strict long-range cSDW order.
We therefore conjecture that the Gutzwiller projection (\ref{Psi_0})
of (\ref{Phi_0}) at half filling is a weak cSDW.
Excluding possible renormalizations,
it has a spin stiffness\cite{chandra_coleman_larkin_90} $\rho_s = S_*^2 J_1$,
with an ordered moment $S_* \sim x_+^{1/2}$. 
Notice that such stripe spin order is consistent with
the product of spin-singlet pairs (\ref{slater_deter_half_fill})
at the limit $x_{\pm}\rightarrow 0$ only if $N/2$ is odd, as required,
in which case the pairs of opposing spins
are separated by the maximum  displacement  ${\bi L}_* = (N/2) a (\hat{\bi x}\pm\hat{\bi y})$.
Two metallic cases exist off half filling. 
The hole Fermi surfaces centered at zero 2D momentum
are larger than the nested ones centered at cSDW momenta
for $\varepsilon_+ < \varepsilon_-$,
where the net concentration of mobile holes at site-orbitals
is proportional to the difference in area between the two: $x = x_+ - x_-$.
The Gutzwiller projection (\ref{Psi_0}) then yields
a variational wavefunction for the $t$-$J$ model (\ref{tJ})
that lies below half filling in this case.
At $\varepsilon_+ > \varepsilon_-$,
on the other hand,
the area of the Fermi surfaces centered at cSDW momenta is larger,
and the Gutzwiller projection (\ref{Psi_0}) now  has a net concentration of
electrons at site-orbitals instead of holes: $-x = x_- - x_+$. 
Each mobile electron must form a spin singlet with a local moment at the site-orbital
in the present infinite-$U_0$ limit.

We now argue for electron-type dispersion of
the low-energy electronic structure 
that emerges at cSDW momenta for Hund's Rule coupling of moderate strength
in the two-orbital $t$-$J$ model (\ref{tJ}) with off-diagonal magnetic frustration.
Let us confine ourselves to the ideal case $t_2^{\perp} = 0$ shown by Fig. \ref{spctrm_cp_a}.
The $3d_{xz}$ and $3d_{yz}$ orbitals
are then good quantum numbers,
which we now label by $k_0 = 0$ and $\pi$, respectively.
The phase factors in 
the one-electron states (\ref{1e_stts}) can then be
set to unity ($\delta_e = 0$),
which means that the one-hole energy (\ref{e_f}) instead becomes
$-\varepsilon_e(k_0,{\bi k}) = \varepsilon_{\parallel}({\bi k})
 + e^{ik_0} \varepsilon_{\perp}({\bi k})$.
Above, the nesting wave vectors
for the $3d_{xz}$ and $3d_{yz}$ orbitals 
are respectively ${\bi Q}_0 = (\pi/a){\hat{\bi y}}$ and ${\bi Q}_{\pi} = (\pi/a){\hat{\bi x}}$.
Assume that the low-energy electronic structure displayed by Fig. \ref{spctrm_cp_a} is
described by a renormalized energy band $\tilde\varepsilon_e(k_0,{\bi k})$
that remains hole-type near ${\bi k}=0$.
In the present case, 
where the $3d_{xz}$ and $3d_{yz}$ orbitals are good quantum numbers,
the dynamical correlation function for true spin fluctuations at cSDW momenta
reduces to the standard Lindhard function (see ref. \cite{graser_09}):
\begin{eqnarray}
{\rm Re}\, \langle {\bi S}(+)\cdot {\bi S}^{\prime}(+)\rangle |_{{\bi Q}_{k_0},\omega}^{\perp} 
\cong  {\pi\hbar^2\over{N_{\rm Fe}}} \sum_{\bi k}^{\qquad\prime}
& n_F(k_0,{\bi k})[1-n_F(k_0,{\bi k}+{\bi Q}_{k_0})]\cdot \nonumber \\
& \cdot(\delta[\omega - \tilde\varepsilon_{e}(k_0,{\bi k}+{\bi Q}_{k_0})+\tilde\varepsilon_{e}(k_0,{\bi k})]\nonumber \\
& -\delta[\omega + \tilde\varepsilon_{e}(k_0,{\bi k}+{\bi Q}_{k_0})-\tilde\varepsilon_{e}(k_0,{\bi k})])
 . \nonumber \\
\label{true_spin_fluct}
\end{eqnarray}
Here, we have used the fact that ${\bi Q}_{k_0} + {\bi Q}_{k_0} = 0$.
Importantly, expression (\ref{true_spin_fluct}) predicts that low-energy cSDW spin fluctuations exist when the
renormalized band $\tilde\varepsilon_{e}(k_0,{\bi k})$
crosses the Fermi level both near ${\bi k}=0$ and near ${\bi k}={\bi Q}_{k_0}$.
Assume next an electron-type dispersion for the nested band of emergent one-particle states.
The spectral weight for such spin fluctuations,
$A(+,{\bi Q}_{k_0}) = \int_0^{\infty} d\omega\, {\rm Re}\, \langle {\bi S}(+)\cdot {\bi S}^{\prime}(+)\rangle |_{{\bi Q}_{k_0},\omega}^{\perp}$,
is then of order $\pi \hbar^2$ 
because of the particle-hole factors in expression (\ref{true_spin_fluct}).
This result is consistent with the present local-moment description (\ref{tJ})
of a cSDW metal with short-range antiferromagnetic order.
On the other hand,
like hidden spin fluctuations (\ref{hidden_spin_fluct}), 
a hole-type dispersion for the emergent band results in a relatively
small spectral weight that instead is equal to $\pi \hbar^2 |x_+ - x_-|$,
where $x_+$ and $x_-$ are the concentrations of holes within the Fermi surface pockets
centered at zero 2D momentum and centered at cSDW momenta, respectively.
It vanishes at half filling, $x_+ = x_-$,
which implies that no cSDW spin fluctuations exist in such case.
In particular,
no spin fluctuations at cSDW momenta
should exist at the quantum-critical point
that separates hidden magnetic order at weak Hund coupling
from cSDW order at strong Hund coupling
in the corresponding  frustrated two-orbital Heisenberg model
over the square lattice\cite{jpr10}.
Also,
no long-range magnetic order of either type exists at the quantum-critical point.
Application of the quantum-fluids analogy for frustrated antiferromagnets\cite{chandra_coleman_larkin_90}
at a quantum critical point
then yields that there exist absolutely no low-energy spin excitations
that carry cSDW momenta there,
neither ``superfluid'' nor  ``normal''.
We believe that this is unlikely.
A comparison of the spectral weights for cSDW  spin fluctuations
therefore argues in favor of electron-type dispersion for the nested band
of one-particle states that emerges at cSDW momenta.

In conclusion, we propose a cSDW metal groundstate,
Eqs. (\ref{Phi_00}), (\ref{Psi_0}) and (\ref{Phi_0}), 
for the two-orbital $t$-$J$ model
with off-diagonal frustration, at moderate Hund's Rule coupling.
It is suggested both by Schwinger-boson-slave-fermion mean-field theory
and by exact results for the low-energy spectrum of one mobile hole.
It notably shows low-energy hidden spin fluctuations at the long-wavelength limit 
because of nesting between the two hole-pockets centered at zero 2D momentum.  
(See Fig. \ref{FS}.)
This type of hidden spin fluctuation may play an important
role in the formation of Cooper pairs in iron-pnictide materials.  
Last, 
the {\it same} arguments can be applied to the hole-pocket Fermi surfaces 
obtained from DFT\cite{singh_du_08}\cite{dong_08}\cite{graser_09},
which are similar to those depicted by Fig. \ref{FS}.
In particular, 
the application of  expression (\ref{Lindhard})
implies that low-energy hidden spin fluctuations must also exist at
the long wavelength limit in such case.

\section{Discussion and Conclusions}
Starting from a local-moment description of a cSDW
over a square lattice of spin-$1$ iron atoms with mobile holes,
we have succeeded in accounting for the nested Fermi surface pockets 
centered at zero 2D momentum and at cSDW momenta 
that  are characteristic of iron-pnictide high-temperature superconductors.
In particular,
zero-energy spin-wave excitations at cSDW momenta combine
with hole Fermi surface pockets centered at zero 2D momentum
to produce Fermi surface pockets centered at cSDW momenta.
The former hole pockets exist because of proximity to
a hidden half metal state with opposing polarized spin over
$3d_{(x+iy)z}$ and $3d_{(x-iy)z}$ orbitals, respectively,
which violates Hund's Rule.
The isotropic $(d+ , d-)$ orbital basis that we choose notably 
maximizes the Hund's Rule coupling (see Appendix A),
and it leads to isotropic Heisenberg exchange coupling constants 
across neighboring spins on the square lattice of iron atoms.
This orbital basis then very likely
minimizes the net magnetic energy
in the two-orbital $t$-$J$ model (\ref{tJ}) at fixed iron moment,
$\langle |\sum_{\alpha} {\bi S}_{i,\alpha}|^2 \rangle^{1/2}$.
Notice that the emergent Fermi-surface nesting
uncovered here is a mirror image of weak-interaction
descriptions for electronic structure in iron-pnictide high-temperature superconductors,
where low-energy spin excitations centered at cSDW momenta are a
{\it result} of Fermi surface nesting\cite{singh_du_08}\cite{dong_08}\cite{graser_09}.
The present local-moment description has also revealed the existence of
hidden low-energy spin excitations between the $d+$ and $d-$ orbitals 
that is intimately tied to nesting between the two hole pockets centered at zero 2D momentum.
DFT calculations of the electronic structure 
in iron-pnictide materials\cite{singh_du_08}\cite{dong_08}\cite{graser_09}
also exhibit
concentric hole-pocket Fermi surfaces similar to those depicted by Fig. \ref{FS}.
We therefore suggest that hidden low-energy spin-fluctuations 
at the long wavelength limit are predicted by DFT as well.

It is useful to contrast our results with those of more ad hoc theoretical models 
that separate local moments from itinerant electrons\cite{Kou_09}\cite{Lv_10}.
Although such models are capable of simultaneously accounting for
the spin-wave spectra and for the Fermi surfaces seen
in iron-pnictide high-temperature superconductors and their parent compounds,
they clearly have less predictive power by virtue of
the explicit separation between the two phenomena.
Early models for iron-pnictide systems that simply add 
Heisenberg exchange interactions and Hund coupling  
to one-electron hopping Hamiltonians that already include nested Fermi surfaces,
but that do not project out double occupancy at an iron site-orbital,
also suffer from this drawback in our opinion\cite{Seo_08}.
In particular, such models essentially operate in the weak-interaction limit,
but they fail to link
Fermi surface nesting to low-energy spin excitations at cSDW momenta
in iron-pnictide high-temperature superconductors.


In summary,
a mean-field theory analysis and an exact diagonalization study
indicate that the two-orbital $t$-$J$ model (\ref{tJ}) for 
iron-pnictide high-temperature superconductors
transits from a cSDW to a hidden half-metal state
with decreasing Hund's Rule coupling
if off-diagonal magnetic frustration exists:
$J_1^{d\pm,d\pm} < J_1^{d\pm,d\mp}$, $J_1^{d\pm,d\mp} > 0$ and $J_2^{\alpha,\beta} > 0$.
Equation (\ref{J_0c}) for the critical Hund coupling 
implies that moving off the QCP by a few percent in the hole concentration
leads to a relatively small deviation:
$|\delta J_{0c} /J_{0c}| \cong 4|t_{\parallel} \delta x / J_{0c}|$.
We propose that the quantum critical point predicted by the mean-field theory
and by exact results (Fig. \ref{spctrm_cp_a})
controls the phase diagram of iron-pnictide high-$T_c$ superconductors
in the vicinity of the transition to the cSDW.
This proposal is consistent with the low-energy spin-excitation spectrum 
and with the low-energy
electronic structure shown by these systems.
In particular,
Schwinger-boson-slave-fermion mean-field theory (\ref{chi_perp}) 
predicts low-energy spinwaves that disperse anisotropically at cSDW momenta,
$(\pi/a){\hat{\bi x}}$ and $(\pi/a){\hat{\bi y}}$,
near the quantum critical point that separates the hidden half metal from the cSDW.
The predicted dispersion of the spinwave spectrum, Fig. \ref{spin_x}, 
notably shows a local maximum
at the N\'eel wavenumber $(\pi/a)({\hat{\bi x}}+{\hat{\bi y}})$,
which agrees with inelastic neutron scattering studies 
of the parent compound CaFe$_2$As$_2$ \cite{zhao_09}.
Further, the anisotropic dispersion that we predict at the QCP for
low-energy spinwaves with cSDW momenta 
is consistent with recent observations of the same
in iron-pnictide superconductors\cite{hayden_10}\cite{Park_10}\cite{Li_10}\cite{liu_12}.
And in agreement with what ARPES reveals for the electronic structure in 
iron-pnictide systems\cite{zabolotnyy_09}\cite{fink09}\cite{brouet09}\cite{zhang11},
both the Schwinger-boson-slave-fermion mean-field theory formulation (\ref{G})
and exact diagonalization of the two-orbital $t$-$J$ model (\ref{tJ})
predict nested Fermi surface pockets around zero 2D momentum and around cSDW momenta 
with purely $3d_{xz}$ and $3d_{yz}$ orbital character at the quantum critical point.
The dispersion of the former is  hole type by construction, 
whereas arguments put forth at the end of the previous section
indicate that the dispersion of the latter is electron type.
Recent ARPES on optimally doped Ba(Fe$_{1-x}$Co$_x$)$_2$As$_2$ notably
find similar electron-pocket Fermi surfaces around cSDW momenta with
purely $3d_{xz}$ and $3d_{yz}$ orbital character\cite{zhang11}.

\begin{figure}
\includegraphics{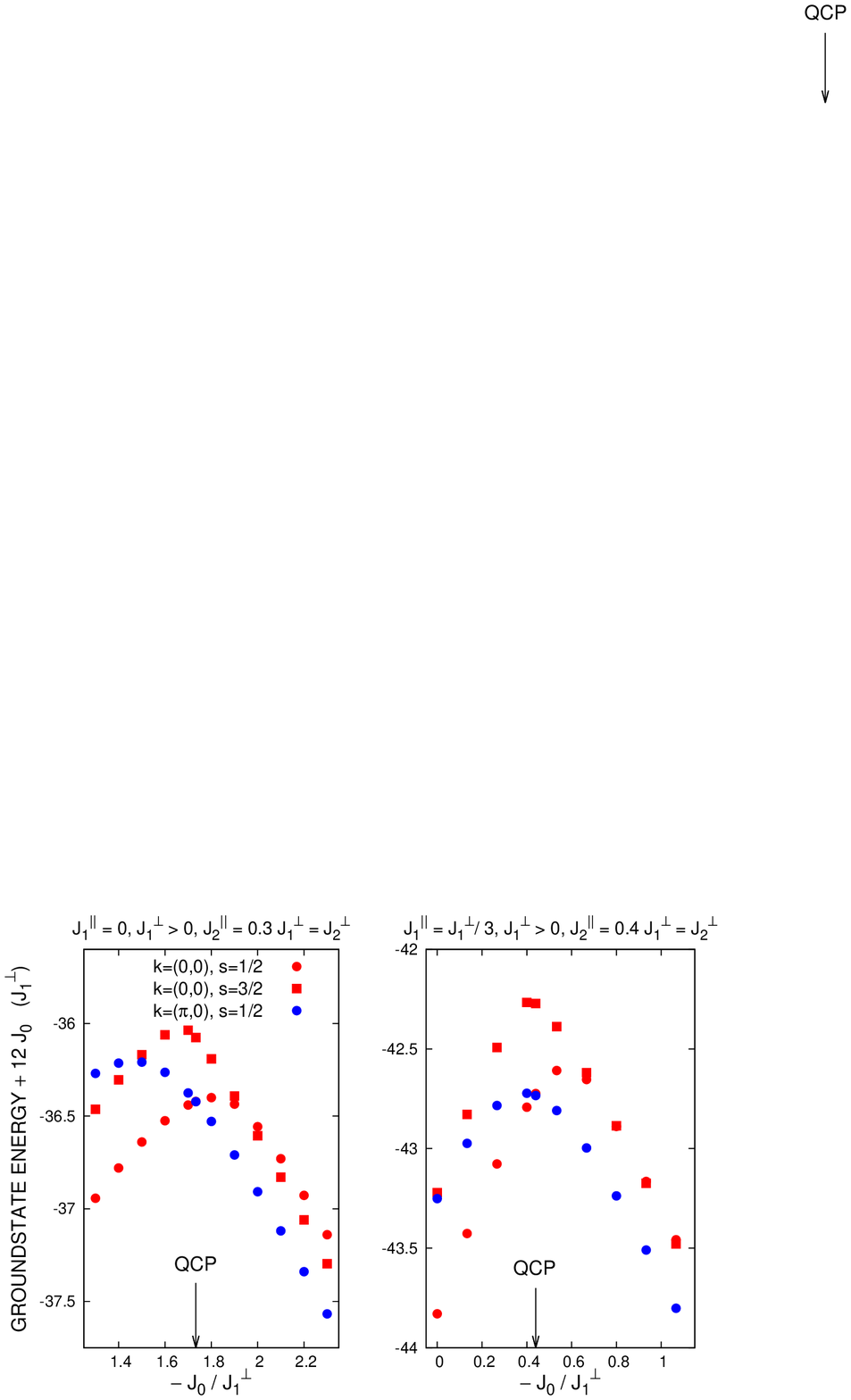}
\caption{How groundstate degeneracy breaks as the Hund's Rule coupling moves off the QCP 
is displayed for one hole 
that roams over a $4\times 4\times 2$ lattice in the 
two-orbital $t$-$J$ model.
Hopping matrix elements are set to
$t_1^{\parallel} = - 5 (J_1^{\parallel} + J_1^{\perp})$,
$t_1^{\perp} ({\hat{\bi x}}) = - 2 (J_1^{\parallel} +  J_1^{\perp})$, 
$t_1^{\perp} ({\hat{\bi y}}) = + 2 (J_1^{\parallel} + J_1^{\perp})$, and
$t_2^{\parallel} = 0 = t_2^{\perp}$.}
\label{e_vs_j0}
\end{figure} 

We finish by pointing out that more recent 
experimental determinations of the electronic structure in
interfacial iron selenides find evidence 
for superconductivity with a record critical temperature\cite{xue_12}\cite{zhou_12}.
In particular, ARPES on a single layer of FeSe 
sees an isotropic superconducting gap at the
electron Fermi surface pockets centered at cSDW momenta
that closes at $T_c = 65$-$70$ K \cite{zhou_13}\cite{peng_13}.
Importantly, ARPES also reveals that the hole bands centered
at zero 2D momentum lie {\it below} the Fermi level in general.  
This experimental observation conflicts with DFT calculations on interfacial iron selenides,
which predict that such hole bands should cross the Fermi surface\cite{B&C}.
We suggest, instead, that interfacial high-$T_c$ superconductivity in iron selenides
is consistent with the quantum critical point found here.  
Figure \ref{e_vs_j0} plots the evolution of the lowest energy levels near the QCP 
as a function of Hund's Rule coupling strength.  
The energies are exact values for the two-orbital $t$-$J$ model (\ref{tJ}) over a
$4\times 4\times 2$ lattice with one mobile hole.  
Next-nearest neighbor hopping is turned off, which yields exact particle-hole symmetry.
Figure \ref{e_vs_j0} therefore also accurately describes the evolution of
the corresponding groundstate energies for
one mobile electron above half filling.
Notice that the groundstate at momentum $(\pi/a)\hat{\bi x}$
lies below the groundstate at zero momentum for Hund's Rule coupling that is stronger
than the critical value.  
The maximum energy difference
between those two states
in the left panel is
approximately $\Delta \varepsilon_{e,h} = J_1^{\perp}/3$
at $J_0 = -1.95\, J_1^{\perp}$.
It is a good fraction of the maximum
dispersion of the low-energy spin-$1/2$ states shown in Fig. \ref{spctrm_cp_a}.
ARPES on mono-layer FeSe superconductors find a 
difference in energy of $\Delta\varepsilon_{e,h} = 15$ meV
between the bottom of the electron band and the top of the hole band\cite{zhou_13}.
Comparing these two yields a lower bound for the Heisenberg exchange coupling constant of order
$J_1^{\perp} \gtrsim  50$ meV. 
Fits of the spin-wave spectrum predicted by the corresponding two-orbital Heisenberg model 
to the dispersion of the spin resonance in Ba(Fe$_{1-x}$Co$_x$)$_2$As$_2$ \cite{hayden_10}
yield a value of $J_1^{\perp} \gtrsim 100$ meV\cite{jpr10}, which lies inside this range.
Last, the right panel in Fig. \ref{e_vs_j0} demonstrates that the emergent
electronic structure at cSDW momenta is robust in the presence of intra-orbital
Heisenberg exchange $J_1^{\parallel} > 0$.
The evolution of the groundstate energy levels
with Hund's Rule coupling described above
therefore potentially accounts for the absence of hole-pocket Fermi surfaces
at zero 2D momentum in single-layer FeSe.

\ack
J.P.R. thanks Ni Ni, Edward Rezayi and Stephan Haas for discussions
and Assa Auerbach for correspondence.
Exact diagonalization of the two-orbital  $t$-$J$
model (\ref{tJ}) was carried out on the SGI Altix 4700
at the 
AFRL DoD Supercomputer Resource Center.
This work was supported in part by the US Air Force
Office of Scientific Research under 
grants nos. FA9550-09-1-0660 and FA9550-13-1-0118 
and by the FCT (Portugal) under grant PTDC/FIS/101126/2008
and grant PEST- OE/FIS/UI0091/2011.

\clearpage

\appendix
\section{}
Here we calculate the Hund's Rule coupling in the 2D subspace spanned by the
$3d_{xz}$ and $3d_{yz}$ orbitals in iron-pnictide materials.
Consider the most general pair of basis states for such orbitals:
\numparts
\begin{eqnarray}
\phi({\bi r}) &= R_{3,2}(r)[(\cos \alpha_0) e^{-i\phi_0} Y_{2,+1}(\Omega)+(\sin \alpha_0) e^{+i\phi_0} Y_{2,-1}(\Omega)],\\
\psi({\bi r}) &= R_{3,2}(r)[-(\sin \alpha_0) e^{-i\phi_0} Y_{2,+1}(\Omega)+(\cos \alpha_0) e^{+i\phi_0} Y_{2,-1}(\Omega)].
\end{eqnarray}
\endnumparts
Notice that $\alpha_0 = 0$ or $\pi/2$ corresponds to the $3d_{(x+iy)z}$/$3d_{(x-iy)z}$ orbital basis,
whereas $\alpha_0 = \pi/4$ and $\phi_0 = 0$ corresponds to the $3d_{xz}$/$3d_{yz}$ orbital basis.
The exchange Coulomb  integral is related to the Hund's Rule exchange coupling constant $J_0$ by
\begin{equation}
-{1\over 2} J_0 = \int d^3r_1 \int d^3r_2\,
 \phi^*({\bi r_1}) \psi({\bi r_1})
{e^2\over{|{\bi r_1}-{\bi r_2}|}} 
 \psi^*({\bi r_2}) \phi({\bi r_2})
\label{ex_int}
\end{equation}
in general.  In the present case, this yields
\begin{eqnarray}
-{1\over 2} J_0 = \int d^3 r_1 &\int d^3 r_2\,
R_{3,2}^2(r_1)
[(\cos \alpha_0)^2 Y_{2,+1}^2(\Omega_1^{\prime})-(\sin \alpha_0)^2 Y_{2,-1}^2(\Omega_1^{\prime})] \nonumber \\
&\cdot {e^2\over{|{\bi r_1}-{\bi r_2}|}} R_{3,2}^2(r_2) 
[(\cos \alpha_0)^2 Y_{2,+1}^{* 2}(\Omega_2^{\prime})-(\sin \alpha_0)^2 Y_{2,-1}^{* 2}(\Omega_2^{\prime})]
\nonumber ,
\end{eqnarray}
where $\Omega^{\prime}$ is the solid angle rotated by $\phi_0$ about the $z$ axis.
The integrals over solid angles $\Omega_1^{\prime}$ and $\Omega_2^{\prime}$ can be performed
in the standard way\cite{B&J} by use of the mathematical identity
\begin{equation}
{1\over{|{\bi r_1}-{\bi r_2}|}} = \sum_{l=0}^{\infty}{r_<^l\over{r_>^{l+1}}} {4\pi\over{2l+1}}
\sum_{m=-l}^l Y_{l,m}(\Omega_1^{\prime}) Y_{l,m}^* (\Omega_2^{\prime})
\label{math_id}
\end{equation}
combined with addition of angular momentum:
\begin{equation}
Y_{2,\pm 1}^2(\Omega) = {1\over 7}\Biggl({15\over{2\pi}}\Biggr)^{1/2} Y_{2,\pm 2}(\Omega) 
+ {1\over 7}\Biggl({10\over{\pi}}\Biggr)^{1/2} Y_{4,\pm 2}(\Omega) .
\end{equation} 
Performing the remaining radial integrals after
substitution of the hydrogenic radial wave function
\begin{equation}
R_{3,2} (r) = {2\over{81}}\Biggl({2\over{15 a_0^3}}\Biggr)^{1/2}\Biggl({r\over{a_0}}\Biggr)^2 e^{-r/3 a_0}
\end{equation}
then yields the final result
\begin{equation}
-{1\over 2} J_0 = {1\over{(2\cdot 3)^{5} (5\cdot 7)^{2}}}
\Biggl[3\, I_2 + 5\Biggl({2\over 3}\Biggr)^2 I_4\Biggr] {e^2\over{a_0}}
(\cos^4\alpha_0 + \sin^4\alpha_0),
\label{J_0}
\end{equation}
where
\begin{equation}
I_4 = 
{12!\over{2^{12}}}\Biggl({2\over 11} - {1\over 12}\Biggl)
\end{equation}
and
\begin{equation}
I_2 =
{12!\over{2^{12}}}\Biggl({2^3\over 9} - 3{2^2\over 10} + 3{2\over 11} - {1\over 12}\Biggl)
\end{equation}
are the strengths of the integrals due to the $l=4$ and $l=2$ channels, respectively.
Note that the binomial-like series that appear above result from the difference
$$(5-l)! - 2^{-(6-l)} s_{11}^{(5-l)} (1/2),$$
where $s_n^{(m)} (x)$ is the $m^{\rm th}$ derivative of the finite geometric series sum
$s_n (x) = \sum_{k=0}^n x^k$.
The Hund's Rule coupling is then
\begin{equation}
-J_0 = {1\over 30.3082} \, {e^2\over 2 a_0}(\cos^4\alpha_0 + \sin^4\alpha_0).
\end{equation}
Notice that $\cos^4\alpha_0 + \sin^4\alpha_0 = 1 - {1\over 2} \sin^2 2\alpha_0$, 
which reaches its maximum value of unity at $\alpha_0 = 0$ or $\pi/2$
and its minimum value of $1/2$ at $\alpha_0 = \pi/4$.
We conclude that the Hund's Rule coupling is largest
in the isotropic $3d_{(x+iy)z}$/$3d_{(x-iy)z}$ orbital basis,
while it is smallest in the anisotropic $3d_{xz}$/$3d_{yz}$ orbital basis from Chemistry.

It is useful to compare the maximum Hund's Rule coupling in 
the isotropic $3d_{(x+iy)z}$/$3d_{(x-iy)z}$ orbital basis
with the on-site Coulomb repulsion in that case:
\begin{equation}
U_0 = \int d^3 r_1 \int d^3 r_2\,R_{3,2}^2(r_1) |Y_{2, 1}(\Omega_1)|^2 
{e^2\over{|{\bi r_1}-{\bi r_2}|}} 
R_{3,2}^2(r_2) |Y_{2, 1}(\Omega_2)|^2.
\end{equation}
Addition of angular momentum yields the identity
\begin{equation}
|Y_{2, 1}(\Omega)|^2 = {1\over{7\pi}} P_0 (\cos\theta) + 
{5\over{28\pi}} P_2 (\cos\theta) - 
{3\over{7\pi}} P_4 (\cos\theta)
\end{equation}
in terms of Legendre polynomials. Substituting it above,
with $P_l(\cos\theta) = [4\pi/(2l + 1)]^{1/2} Y_{l,0} (\Omega)$,
along with the mathematical identity (\ref{math_id}), yields the result
\begin{equation}
U_0 = {1\over{2^{6} 3^{5} 5^{2}}}
\Biggl[I_0 + {1\over{7^2}} I_2 + \Biggl({4\over{3\cdot 7}}\Biggl)^2 I_4\Biggr] {e^2\over{a_0}}
\label{Hubbard_U}
\end{equation}
for the Coulomb integral, where
\begin{equation}
I_0 =
{12!\over{2^{12}}}\Biggl({2^5\over 7} - 5 {2^4\over 8} + 10 {2^3\over 9} -
10 {2^2\over 10} + 5 {2\over 11} - {1\over 12}\Biggl)
\end{equation}
is the strength of the integral in the $l=0$ channel.
The ratio of the on-site Coulomb repulsion (\ref{Hubbard_U})
to the Hund's Rule coupling (\ref{J_0}) is then
\begin{equation}
-{2U_0\over{J_0}} = 10.6743 .
\label{ratio}
\end{equation}
Study of the exchange integral (\ref{ex_int}) yields that this ratio coincides with the
ratio $J_1^{d+, d+}({\rm drct}) / J_1^{d+,d-}({\rm drct})$ in the regime $a_0\gg a$.

\clearpage

\section{}
Here, we evaluate the Gutzwiller projection (\ref{Psi_0}) of 
the ``filled'' Fermi gas (\ref{Phi_00}).
Specifically, the Fermi level $\varepsilon_+$ (and $\varepsilon_-$)
lies above the maximum of both bands,
$\varepsilon_e(0,{\bi k})$ and $\varepsilon_e(\pi,{\bi k})$, at ${\bi k} = 0$.
The Fermi gas (\ref{Phi_00}) is then a Slater determinant over all of the allowed
one-electron states restricted to ``black squares'' of the ``checkerboard'' in 2D momentum.
It can be written in the form
\begin{eqnarray}
\Phi_0(1,\ldots ,{\cal N}) = ({\cal N}!)^{-1/2} 
\sum_{[q]} \sum_{p\in [q]} ({\rm sgn}\, p)
&\Biggl[\prod_{m=1}^{N_{\rm Fe}} \phi_{k_m}(p(m)) \uparrow_{p(m)}\Biggr]\cdot
\nonumber \\
&\cdot\Biggl[\prod_{n=1+N_{\rm Fe}}^{\cal N} \phi_{k_{n^{\prime}}}(p(n)) \downarrow_{p(n)}\Biggr] ,
\nonumber \\
\label{slater_deter}
\end{eqnarray}
where $n^{\prime} = n-N_{\rm Fe}$,
and where $[q]$ denote equivalence classes of permutations $p$ that do not flip any spins.
In particular, we can write 
$p = q p_{\uparrow} p_{\downarrow}$,  where $p_{\uparrow}$ and $p_{\downarrow}$ denote
permutations of $1,2,\ldots, N_{\rm Fe}$ and of $1+N_{\rm Fe},2+N_{\rm Fe},\ldots,{\cal N}$
respectively.
Above, $k_1, k_2, \ldots, k_{N_{\rm Fe}}$ is 
a list of {\it all} 3-momenta on the ``black squares''.
The Slater determinant (\ref{slater_deter}) can then be written more explicitly as
\begin{eqnarray}
\Phi_0(1,\ldots ,{\cal N}) = ({\cal N}!)^{-1/2} 
&\sum_{[q]} ({\rm sgn}\, q) \cdot \nonumber \\
&\cdot\Biggl[\sum_{p_{\uparrow}} ({\rm sgn}\, p_{\uparrow})
\prod_{m=1}^{N_{\rm Fe}} \phi_{k_m}(q p_{\uparrow}(m)) \uparrow_{q p_{\uparrow}(m)}\Biggr]\cdot
\nonumber \\
&\cdot\Biggl[\sum_{p_{\downarrow}} ({\rm sgn}\, p_{\downarrow})
\prod_{n=1+N_{\rm Fe}}^{\cal N} \phi_{k_{n^{\prime}}}(q p_{\downarrow}(n)) \downarrow_{q p_{\downarrow}(n)}\Biggr] .
\nonumber \\
\label{slater_deter_explct}
\end{eqnarray}
Application of the mathematical identity ${\rm det}\, AB = ({\rm det}\, A) ({\rm det}\, B)$
yields
${\rm det}\, \phi_{k_m}(n) = 
e^{i\phi_0} {\rm det}\, \delta_{\alpha_m,\beta_n}(\delta_{i_m,j_n} + \delta_{i_m^*,j_n})/2^{1/2}$,
where $e^{i\phi_0} = {\rm det}\, 2^{1/2} \phi_{k_m}(\alpha_n,i_n)$.  
Here $(\alpha_1,i_1), (\alpha_2, i_2), \ldots , (\alpha_{N_{\rm Fe}}, i_{N_{\rm Fe}})$
is a list of all orbitals and of 
all sites within a single subsquare in Fig. \ref{split_square}.
On the other hand, $\beta_n$ and $j_n$ represent the orbital and 
the site of electron $n$ with spin $\uparrow (\downarrow )$.
The last determinant lies on the unit circle of the complex plane because 
its argument is an $N_{\rm Fe} \times N_{\rm Fe}$ unitary matrix.
Substituting the previous identity in for the spin-up and for the spin-down determinants
in expression (\ref{slater_deter_explct}) yields
the following expression for the Slater determinant up to a phase factor:
\begin{eqnarray}
\langle 1,\ldots ,{\cal N}|\Phi_0\rangle
= & ({\cal N}!)^{-1/2} 
\sum_{p}  ({\rm sgn}\, p) \cdot \nonumber \\
& \cdot \Biggl[\prod_{m=1}^{N_{\rm Fe}} 2^{-1/2} (\langle p(m)|\alpha_m,i_m\rangle +
\langle p(m)|\alpha_m,i_m^*\rangle) \uparrow_{p(m)}\Biggr]\cdot
\nonumber \\
&\cdot\Biggl[\prod_{n=1+N_{\rm Fe}}^{\cal N} 2^{-1/2} (\langle p(n)|\alpha_{n^{\prime}},i_{n^{\prime}}\rangle +
\langle p(n)|\alpha_{n^{\prime}},i_{n^{\prime}}^*\rangle) \downarrow_{p(n)}\Biggr] .
\nonumber \\
\label{slater_deter_intrmdt}
\end{eqnarray}
%
After grouping together common site-orbital factors into spin-singlet pairs, 
and subsequently projecting out double occupancy per site-orbital, 
we obtain the final expression for the Gutzwiller wavefunction:
\begin{eqnarray}
\langle 1,\ldots,{\cal N}|\Psi_0\rangle = 
{1\over{\sqrt{2^{N_{\rm Fe}}{\cal N}!}}} \sum_p ({\rm sgn}\, p) \prod_{m=1}^{N_{\rm Fe}}
& {1\over \sqrt 2}
(\uparrow_{p(m)}\downarrow_{p(m^{\prime})}-\downarrow_{p(m)}\uparrow_{p(m^{\prime})})\cdot \nonumber \\
& \cdot \langle p(m), p(m^{\prime})|\alpha_m,i_m; \alpha_m,i_m^*\rangle_+  , \nonumber \\
\label{slater_deter_final}
\end{eqnarray}
where $m^{\prime} = m + N_{\rm Fe}$, and where
$|\alpha,i; \beta,j\rangle_+ = 
(|\alpha,i\rangle |\beta,j\rangle + 
|\beta,j\rangle |\alpha,i\rangle)/\sqrt 2$.
It describes a featureless paramagnetic insulator composed of a product of spin-singlet pairs
within the same $d+$ and $d-$ orbitals that entangle opposite spins
separated by the maximum displacement ${\bi L}_*$.

A hole excitation about the above spin-singlet paramagnetic insulator with 2D momentum on the
``black squares'' of the ``checkerboard'' is then just the Gutzwiller projection (\ref{Psi_0})
of the ``filled'' band (\ref{Phi_00}) with that state missing.  
To obtain a hole excitation
that has momentum on the ``white squares'', 
we repeat the previous calculation of the ``filled'' band case, 
but with the 2D momenta restricted to the ``white squares'':  
${\bi k} = (2\pi n_x / N a, 2\pi n_y / N a)$,
where $n_x$ and $n_y$ are respectively even and odd integers,
or vice versa.
The one-electron states
are then antiperiodic on a tilted subsquare in Fig. \ref{split_square}.
In particular, 
the relative  plus sign between sites separated by the maximum displacement ${\bi L}_*$
in the 
one-electron states
that appears in Eqs. (\ref{1e_stts})  and (\ref{slater_deter_intrmdt}) 
is replaced by a relative minus sign in this case.
The Gutzwiller projection (\ref{slater_deter_final}) remains unchanged, however.
A hole excitation with 2D momentum on the ``white squares'' then is simply the Gutzwiller
projection (\ref{Psi_0}) of the ``filled'' band (\ref{Phi_00}) over the ``white squares'',
but with that state missing.

\clearpage


\end{document}